\pdfoutput=1

\documentclass{statsoc}
\usepackage{numprint}
\usepackage{amsmath}
\usepackage{graphicx}
\usepackage{tabularx}
\usepackage{subfigure}
\usepackage{float}
\usepackage{longtable}
\usepackage{url} 
\usepackage{color}
\usepackage{verbatim}
\usepackage{placeins}
\usepackage{multirow}
\usepackage{verbatim}
\usepackage{geometry}

\newcommand{\hl}[1]{#1}
\newcommand{\rev}[1]{#1}
\newcommand\numberthis{\addtocounter{equation}{1}\tag{\theequation}}

\title[Hierarchical modelling of genetic interaction]{Bayesian hierarchical modelling for inferring genetic interactions in yeast}
\author[Jonathan Heydari, Conor Lawless, David A.\ Lydall and Darren J.\ Wilkinson]{Jonathan Heydari, Conor Lawless, David A.\ Lydall and Darren J.\ Wilkinson}
\address{Newcastle University, UK}
\email{d.j.wilkinson@ncl.ac.uk}
\begin{document}
 \begin{abstract}
Quantitative Fitness Analysis (QFA) is a high-throughput experimental and computational methodology for measuring the growth of microbial populations.  
QFA screens can be used to compare the health of cell populations with and without a mutation in a query gene in order to infer genetic interaction strengths genome-wide, examining thousands of separate genotypes. 
We introduce Bayesian, hierarchical models of population growth rates and genetic interactions that better reflect QFA experimental design than current approaches.
Our new approach models population dynamics and genetic interaction simultaneously, thereby avoiding passing information between models via a univariate fitness summary.
Matching experimental structure more closely, Bayesian hierarchical approaches use data more efficiently and find new evidence for genes which interact with yeast telomeres within a published dataset.
\end{abstract}

 \section{\label{sec:Introduction}Introduction}

There are many reasons to study the growth of microbes, including to prevent the growth of pathogenic bacteria or fungi and to encourage the growth of yeasts in industrial applications or during food production.
Another reason is the study of eukaryotic microbes, such as the yeasts \emph{Saccharomyces cerevisiae} and \emph{Schizosaccharomyces pombe}, as biological models of cells in higher eukaryotes (e.g. of human cells).

Evolutionary fitness in a given environment: the probability of genetic material from an individual contributing to the gene pool of the next generation, is an important characteristic of a population that is optimised by natural selection.
Rate of cell division is a major component of fitness, directly affecting the ability of individuals to compete for resources such as space and nutrients.
By measuring and comparing the growth rates of microbial populations (cultures) we can assess and rank the fitness or health of such populations in a given environment or in a given genetic background.

Quantitative Fitness Analysis (QFA) is a method for measuring the growth and fitness of independent microbial cultures inoculated onto solid agar surfaces \citep{jove, QFA1}.
During QFA we inoculate cell cultures at densities of between 96 and 1536 cultures per plate of agar, repeatedly photographing cultures as they grow, converting photographs to quantitative estimates of cell density \citep{Colonyzer}.
We summarise observations of increasing cell density with time (growth curves) by fitting population growth models to observed data.
We use fitted model parameters, such as the intrinsic growth rate parameter of the logistic growth model, to define several measures of culture fitness \citep{QFA1}.

Quantifying the fitness of hundreds of strains on a single plate, under identical environmental conditions, allows a range of powerful experimental designs.
Biological experiments examining the effect of a condition on selected populations, or the effect of selected conditions on one population, are often called screens.
Screening independent replicate cultures with the same genotype allows us to measure biological heterogeneity and to capture technical error (which represents the effect of measurement error, fungal and bacterial contamination, positioning errors and agar cracking in these experiments).
Comparing cultures with different genotypes allows us to explore the relative importance of genes and gene products in a given environment or genetic background.
An important reason for carrying out QFA is to compare the fitnesses of cultures with distinct genotypes in order to quantify the strength of interaction between genes (epistasis).
Screening fitnesses and genetic interaction strengths on a genome-wide scale allows us to study the behaviour of gene products in living cells systematically.
Ready-made genome-wide libraries of strains with distinct genotypes (each with an individual gene deleted, for example) are available and can be mated with selected strains to generate libraries targeted at particular biological processes of interest.
A typical high-throughput, genome-wide QFA screen, examining the fitness of replicate cultures of ~5,000 of different genotypes, includes hundreds of plates that are inoculated, photographed and incubated by laboratory robots.

The genome-wide QFA experiments that we re-analyse in this paper (see Section 4) were designed to inform us about telomere biology in eukaryotic cells.
Telomeres are the ends of linear chromosomes found in most eukaryotic organisms \citep{greider1985identification}, capping chromosome ends to ensure genetic stability, and are usually required for cells to progress through the cell cycle. 
Functional telomere caps help to prevent cancer and, since human telomeres shorten at each round of cell division \citep{telo}, some researchers claim that telomere-induced replicative senescence is an important component of human ageing.
QFA experiments were carried out using \emph{Saccharomyces cerevisiae} (brewer's yeast), a model eukaryotic organism widely used to study genetics.
Yeasts are ideal for genome-wide analysis of gene function, as genetic modification of yeast cells is relatively straightforward and yeast cultures grow quickly; millions of yeast cells can be grown overnight, whereas the same number of human cells could take weeks to grow.

In these experiments, we used a genome-wide collection of \emph{S. cerevisiae} strains, each carrying one of the set of about 5,000 single Open Reading Frame (ORF) deletions that are not essential for cell survival. 
An open reading frame is a DNA sequence containing no stop codons, which means that it has the potential to be translated into a protein or peptide.
We refer to the mutations in this collection as $\emph{orf}\Delta$s; $\Delta$ is the standard genetics nomenclature for a deletion.
Identifying ORFs from sequences is the first step in identifying genes, and using a library of ORFs allows the possibility of discovering biological function for sequences that were previously thought to be untranslated.
However, the majority of ORFs in the collection we analyse have been confirmed as genes of known function and so $\emph{orf}\Delta$s are largely equivalent to gene deletions.

The strain collection was mated with a (query) background strain carrying the \emph{cdc13-1} mutation, chosen for its relevance to telomere biology, to give a new library of strains carrying two mutations.
Comparing fitnesses with a second, new library of strains, built from the deletion collection mated with a strain carrying a neutral control background mutation (\emph{ura3}$\Delta$) allows the separation of the effect of the \emph{cdc13-1} mutation from that of deletions from the original collection.

More generally, we use QFA to infer genetic interaction strengths by comparing fitnesses in two QFA screens: a control screen and a query screen.
All strains within a query screen differ from their control screen counterparts by a common condition such as a background gene mutation, drug treatment, temperature or other treatment.
To identify strains that interact with the query condition we can compare the corresponding fitness responses for each strain in the library under the query and control conditions.
Interactions with the query condition are identified by finding gene disruptions in the query screen whose fitnesses deviate significantly from those predicted by a theoretical model of genetic independence, given the fitness of corresponding gene disruptions in the control screen.
Independent replicate cultures are inoculated and grown across several agar plates for each strain under each condition to capture biological heterogeneity and measurement error.  

In the original analysis presented by \citet{QFA1}, logistic models of population growth were fitted to observed cell density time courses by least squares, thereby generating a univariate fitness estimate for each time course.
A linear model, predicting query strain fitness given control strain fitness, consistent with Fisher's multiplicative model of genetic independence, was used to test for genetic interaction between the query mutation and each deletion from the deletion collection. 
The significance of observed interactions was assigned using a simple frequentist linear modelling approach. 
A major limitation of the statistical model used in \cite{QFA1} is that it assumes that replicate culture fitness variances are the same for each $\emph{orf}\Delta$. 
We expect that explicit modelling of heterogeneity will allow more robust identification of interactions, particularly where variability for a particular strain is unusually high (e.g. due to experimental difficulties).

Other large-scale quantitative genetic interaction screening approaches exist, such as E-MAP \citep{emap} and SGA \citep{sgaboone}, but we expect QFA to provide higher quality fitness estimates by using a culture inoculation technique which results in a wider range of cell densities during culture growth  and by capturing complete growth curves instead of using single time point assays \citep{Colonyzer}. 
QFA as presented by \citet{QFA1} and alternative genetic interaction screening approaches mentioned above use frequentist statistical methods that cannot account for all sources of experimental variation and do not partition variation into population, genotype and repeat levels.
Further, the frequentist statistical approaches used in the methods above cannot incorporate prior beliefs.

With the Bayesian approach \citep{Bayth} that we adopt in this paper, we have more flexibility of model choice, allowing us to match model structure more closely to experimental design.
Bayesian analysis allows us to use binary indicators to describe the evidence that each $\emph{orf}\Delta$ interacts with the query mutation in terms of probability. 
Currently there is no standard frequentist approach which can deal with inference for a hierarchical model that simultaneously models logistic growth parameters and probability of genetic interaction. 
Using Bayesian hierarchical modelling \citep{Zhang2014, GelmanMultilevel} we look to extract as much information as possible from valuable QFA datasets.

Following the approach for determining epistasis from the comparison of two QFA screens presented by \citet{QFA1}, we developed a two-stage approach to this problem: 
$i)$~a hierarchical logistic growth curve model is fitted to cell density measurements to estimate fitness, then $ii)$~fitness estimates are input to a hierarchical interaction model.
Next, we developed a unified approach which we refer to as the joint hierarchical model (JHM). 
The JHM models mutant strain fitnesses and genetic interactions simultaneously, without having to pass information between two separate models.
The JHM can also allow two important, distinct, microbial fitness phenotypes (population growth rate and carrying capacity) to provide evidence for genetic interaction simultaneously.

The paper is organized as follows: Section~\ref{Defining_Fitness} describes the data from a typical QFA experiment.
The two new models for Bayesian QFA are outlined in Section~\ref{BayInf}.
In Section~\ref{Application_data} the new Bayesian models are applied to a previously analysed QFA data set for identifying yeast genes interacting with a telomere defect. 
Section~\ref{sec:Discussion} discusses the relative merits of the newly developed Bayesian methods. 
 
  \section{\label{Defining_Fitness}Defining Fitness}
\begin{figure}[h!]
  \centering
\includegraphics[width=10cm]{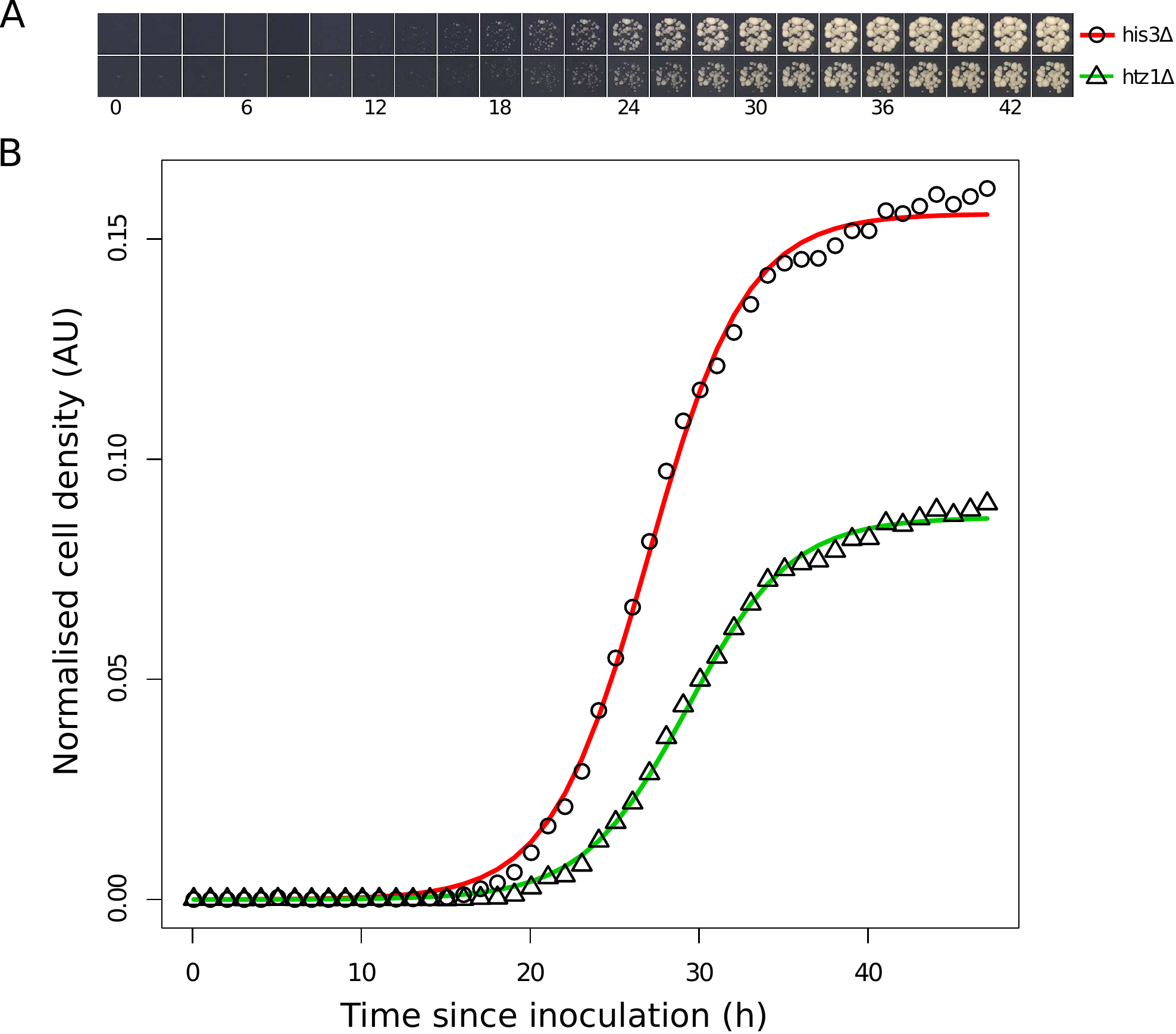}
\caption{
QFA image data and growth curves.
A) Timelapse images for two genetically modified \emph{S. cerevisiae} cultures with different genotypes (indicated) corresponding to the time series measurements plotted in panel B.
B) Timecourse cell density estimates derived from analysis of the timelapse images in panel A together with (least squares) fitted logistic growth curves.
}
\label{fig:spot2}
\end{figure}

Observing changes in cell number in a microbial culture is the most direct way to estimate culture growth rate, an important component of microbial culture fitness.
Direct counting of cells in a high-throughput experiment is not practical and so, during QFA, cell density estimates are made instead from culture photographs.
Robotic assistance is required for both culture inoculation and image capture during genome-wide screens which can include approximately 5,000 independent genotypes.
We use estimates of the integrated optical density (IOD) generated by the image analysis tool Colonyzer \citep{Colonyzer} to capture cell density dynamics in independent cultures during QFA (see Figure~\ref{fig:spot2}A).

Density estimates, scaled to normalise for camera resolution, are gathered for each culture and a dynamic model of population growth, the logistic model $\dot{x}=rx(1 - x/K)$ \citep{Verhulst1847}, is fit to the data. 
The logistic model ODE has three parameters: $K$, $P$ and $r$, the carrying capacity (maximum achievable population density), culture inoculum density (initial condition) and culture growth rate respectively, and has the following analytical solution:
\begin{equation}
  x(t;\theta)=\frac{K P e^{rt}}{K + P \left( e^{rt} - 1\right)} \text{, where } P=x(0) \text{ and } \theta=(K,r,P). 
  \label{eq:logistic}
  \end{equation}
This model describes self-limiting populations undergoing approximately exponential growth which slows as population density increases.
During QFA, self-limited growth occurs because nutrients found in the solid agar substrate are consumed by the growing cell population.
Ultimately the population density saturates at the carrying capacity once available nutrients are exhausted (see Figure~\ref{fig:spot2}).

We can construct several distinct, quantitative fitness measures based on fitted logistic model parameters. 
\cite{QFA1} present three univariate measures suitable for QFA: Maximum Doubling Rate $(D_R)$ and Maximum Doubling Potential $(D_P)$, and their product $D_R\times D_P$,
  \begin{equation}
  \label{eq:MDRMDP}
 \text{where{\qquad}} D_R=\frac{r}{\log\left(2\frac{K-P}{K-2P}\right)}\text{{\qquad}and\qquad} D_P=\frac{\log\left(\frac{K}{P}\right)}{\log(2)}. 
  \end{equation}
$D_R$ captures the rate at which microbes divide immediately after inoculation, when experiencing minimal intercellular competition or nutrient stress.
A strain's growth rate largely dictates its ability to outcompete any neighbouring strains.
$D_P$ captures the number of divisions the culture is observed to undergo before saturation.
A strain which can divide more often than its neighbours in a specific environment also has a competitive advantage.

The choice of a single overall fitness score depends on the aspects of microbial physiology most relevant to the biological question at hand.
Typically the fitness definition $D_R\times D_P$ is used in QFA to account for both attributes simultaneously.

\subsection{\label{sec:a_labelb}Epistasis}
Epistasis is the phenomenon where the effects of one gene are modified by those of one or several other genes \citep{epis4}.
As presented in \cite{QFA1}, here we use Fisher's multiplicative model of genetic independence \citep{cordell2002epistasis,epis1} to represent the expected relationship between control strain fitness phenotypes and those of equivalent query strains in the absence of genetic interaction. 
We interpret genotypes for which the query strain fitness deviates significantly from this model of genetic independence as interacting significantly with the query mutation.
Here, we use square bracket notation to represent a quantitative fitness measure. 
For example $[wt]$ and $[query]$ represent wild-type and query mutation fitnesses respectively.
$\emph{orf}\Delta$ is standard genetics nomenclature for the genotype of a strain with single gene ($\emph{orf}$) deleted.
We use this standard nomenclature to refer to an arbitrary strain from the deletion collection. 
We define new nomenclature to describe a strain containing two mutations.
For example, $query:\emph{orf}\Delta$ represents a strain with the query mutation along with an arbitrary single gene deletion.
We use this nomenclature to refer to an arbitrary strain from the new query strain library constructed by crossing or mating a strain containing the query mutation with each of the strains in the genome-wide deletion collection.
Fisher's multiplicative model of genetic independence can be written as follows:
\begin{eqnarray}
\label{eq:epistasis}
[query:\emph{orf}\Delta]\times [wt] &=& [query]\times [\emph{orf}\Delta]\\
\Rightarrow [query:\emph{orf}\Delta] &=& \frac{[query]}{[wt]}\times [\emph{orf}\Delta]. \label{eq:linear}
\end{eqnarray}

In (\ref{eq:linear}),~$\frac{[query]}{[wt]}$ is a constant for a given pair of QFA screens, meaning that if this model holds, there should be a linear dependence between $[query:\emph{orf}\Delta]$ and $[\emph{orf}\Delta]$ for all deletions $\emph{orf}\Delta$. 
During genome-wide screens of thousands of independent $\emph{orf}\Delta$s we can assume that the majority of gene mutations in the library do not interact with the chosen query mutations.  
Therefore, even if the query or wild-type fitnesses are not available to us, we can still estimate the slope of this linear model by fitting it to all available fitness observations, before testing for strains which deviate significantly from the linear model. 
Any extra background condition, such as a gene mutation common to both the control and query strains (e.g. triple instead of double deletion strains for the query and control data sets), may change the biological interpretation of the interaction, but the same linear relationship is applicable.
Besides the multiplicative model, there are other definitions for epistasis such as additive, minimum and log \citep{epis2}. 
Minimum is a suboptimal approach which may allow ``masking'' of interactions \citep{epis2}.
In this paper, we use a multiplicative interaction model (\ref{eq:epistasis}), but we note that this is equivalent to \hl{an additive} interaction model when looking at fitnesses on the log scale \citep{epis3}.  
Multiplicative and additive models are equivalent provided fitness data are scaled appropriately \citep{cordell2002epistasis}.

\subsection{\label{sec:previous} Previous QFA methodology}

\cite{QFA1} present QFA where the logistic growth model (\ref{eq:logistic}) is fit to experimental data by least squares to give parameter estimates $(\hat{K},\hat{r})$ for each culture time course (each $\emph{orf}\Delta$ replicate).  
Inoculum density is assumed known and the same across all $\emph{orf}\Delta$s and their repeats.
\hl{After inoculating approximately 100 cells per culture, during the first several cell divisions there are so few cells that culture cell densities remain well below the detection threshold of cameras used for image capture and so, without sharing information across all $\emph{orf}\Delta$ repeats, $P$ cannot be estimated directly.}
\rev{It is therefore necessary to fix $P$ to the same value for both screens, using an average estimate of $P$ from preliminary least squares logistic growth model fits.} 
Fitting the model to each $\emph{orf}\Delta$ repeat separately means there is no sharing of information within an $\emph{orf}\Delta$ or between $\emph{orf}\Delta$s when determining $\hat{K}$ and $\hat{r}$. 

Quantitative fitness scores ($F_{cm} = D_{R,cm}\times D_{P,cm}$) for each culture were defined (see (\ref{eq:MDRMDP}) for definitions of $D_R$ and $D_P$). 
The index $c$ identifies the condition for a given $\emph{orf}\Delta$: $c=0$ for the control strain and $c=1$ for the query strain. 
$m$~identifies an $\emph{orf}\Delta$ replicate.
Scaled fitness measures $\tilde{F}_{cm}$ are calculated for both the control and query screen such that the mean across all $\emph{orf}\Delta$s for a given screen is equal to 1.
After scaling, any evidence that $\tilde{F}_{0m}$ and $\tilde{F}_{1m}$ are significantly different will be evidence of genetic interaction.

The linear model 

\begin{align}
  \label{eq:lm}
		\begin{split}
	\tilde{F}_{cm} &= \mu+\gamma_{c}+\varepsilon_{cm}, \text{ where $\gamma_{0}=0$}\\
	\varepsilon_{cm} &\sim \operatorname{N}(0,\sigma^{2}), \text{ where $ \varepsilon_{cm}$ is i.i.d.}
		\end{split}
\end{align}
was fitted to the control and query strain scaled fitness measure pairs for all unique $\emph{orf}\Delta$s in the gene deletion library. 
In (\ref{eq:lm}), $\gamma_{1}$ represents the estimated strength of genetic interaction between the control and query strain.
If the scaled fitnesses for the control and query strain are equivalent
for a particular \emph{orf}$\Delta$ such that they are both estimated by some $\mu$, i.e. no evidence of genetic interaction, we would expect $\gamma_c = 0$.
The model was fit by maximum likelihood, using the R function ``lmList'' \citep{nlme} with variation assumed to be the same for all strains in a given screen and the same for both control and query screens.
Hence, for every gene deletion from the library an estimate of $\gamma_{1}$ was generated together with a p-value for whether it was significantly different from zero.

False discovery rate (FDR) corrected q-values were then calculated to determine levels of significance for each $\emph{orf}\Delta$. 
\rev{\citet{QFA1} use the Benjamini-Hochberg test \citep{ben_hoc} for FDR correction.
This test is commonly used in genomic analyses as although it assumes independence of test statistics, even if positive correlation exists between tests; the result is that FDR estimates are slightly conservative.}
Finally a list of $\emph{orf}\Delta$ names, ranked by q-values, was output and $\emph{orf}\Delta$s with q-values below a significance cut-off of 0.05 were classed as showing significant levels of genetic interaction with the query mutation.

\subsection{\label{sec:remodel} Random Effects Model}

We attempted to improve on the \cite{QFA1} modelling approach within the frequentist paradigm by accounting for the hierarchical structure of the data with a random effects model \citep{nlme} of genetic interaction:

\begin{align*}
f_{clm}&= \mu_c+Z_l+\gamma_{cl}+\varepsilon_{clm} \\
\mu_{c}&=\begin{cases}
\mu+\alpha  & \text{if } c=0; \\
\mu & \text{if } c=1,
\end{cases}\qquad
&\gamma_{cl}&=\begin{cases}
0  & \text{if } c=0; \\
\gamma_{l} & \text{if } c=1,
\end{cases}\\
Z_l&\sim \mathcal{N}(0,{\sigma_Z}^2)
&\varepsilon_{clm} &\sim \mathcal{N}(0,\sigma^2). \numberthis \label{REMeqs}
\end{align*}

In the random effects model (REM) (\ref{REMeqs}) and in models presented below, $c$~identifies the condition for a given $\emph{orf}\Delta$, $l$~identifies a particular $\emph{orf}\Delta$ from the gene deletion library and $m$ identifies a repeat for a given $\emph{orf}\Delta$.
In (\ref{REMeqs}) we use previously estimated $F_{cm}$ to quantify interaction for all $\emph{orf}\Delta$s simultaneously.
Introducing a random effect $Z_l$ allows us to account for between subject variation by estimating a single ${\sigma_Z}^2$.
Unlike the \cite{QFA1} approach, we do not scale the observed values ${F}_{clm}$ and instead introduce a parameter to model a condition effect $\mu_c$.
$\gamma_{cl}$ represents the estimated strength of genetic interaction between an $\emph{orf}\Delta$ and our query mutation. 
For a multiplicative model of epistasis we use an additive model to describe log transformed data $f_{clm}=\log(F_{clm}+1)$, where ${F}_{clm}$ are our observed fitnesses.
We use the Benjamini-Hochberg test to correct for multiple testing in order to make a fair comparison with the \citet{QFA1} approach.

We find that $\emph{orf}\Delta$ level variation in fitness cannot be modelled efficiently as random effects under the frequentist paradigm, which forces us to assume constant variance for all $\emph{orf}\Delta$s.  
The large number of random effects required (control and query observations for each of about 5,000 $\emph{orf}\Delta$s in a genome-wide screen) to model variances at the $\emph{orf}\Delta$ level resulted in inference involving large matrix computations that either took too long to complete or were not possible using the computing hardware available to us.  
Similarly we found that it is not practical to model genetic interaction and cell population growth curves simultaneously as random effects under the frequentist paradigm.
We attempted to model repeat level variation with a Normal distribution by fitting a model with a log-link function; however none of the non-linear model maximum likelihood algorithms we tried converged.
 
 \section{\label{BayInf}Bayesian hierarchical model inference}
As an alternative to the maximum likelihood approach presented by \cite{QFA1} and the REM, we present a Bayesian, hierarchical methodology where \emph{a priori} uncertainty about each parameter value is described by probability distributions \citep{Bayth} and information about parameter distributions is shared across $\emph{orf}\Delta$s and conditions.
Plausible frequentist estimates from across 10 independent, unpublished QFA data sets, \hl{including a wide range of different background mutations and treatments} were summarised to establish and quantify our \emph{a priori} uncertainty in model parameters. 

\rev{First and foremost, prior distributions describe our beliefs about parameter values.  
Priors should be at least diffuse enough to capture all plausible values (to capture the full range of observations in the datasets) and at least restrictive enough to rule out physically implausible values (to ensure efficient inference).
Priors that are excessively vague are not consistent with the Bayesian paradigm and if they are unnecessarily diffuse can also result in computational difficulties during inference (see below for further details).
The computational time required to overcome mixing problems from careless choice of prior distributions are likely to be considerable when fitting a large, hierarchical model to a rich dataset.
Although using conjugate priors would allow slightly faster inference, we find that, for this particular application, the conjugate priors available for variance parameters \citep{Gelmanprior} are either too restrictive at low variance (Inverse-gamma), not restrictive enough at low variance (half-t family of prior distributions) or are non-informative or largely discard the prior information available (Uniform).
Here we have chosen the non-conjugate Log-normal distribution as a prior for precision parameters as we find that when appropriately parameterised the distribution reflects our prior beliefs about precision parameters and is only restrictive at extremely high and low variances.}

\rev{We use three types of distribution to model parameter uncertainty: Log-normal, Normal and scaled t-distribution with three degrees of freedom.
Particular care is needed in the choice of distributions for parameters which are in some sense close to the data, in order to ensure that the model is flexible enough to describe high-resolution datasets such as those captured during QFA.
We use the Log-normal distribution to describe parameters which are required to be non-negative (e.g. parameters describing precisions, or repeat-level fitnesses) or parameter distributions which are found by visual inspection to be asymmetric.  We use the Normal distribution to describe parameters which are symmetrically distributed (e.g. some prior distributions and the measurement error model) and we use the t-distribution to describe parameters whose uncertainty distribution is long-tailed (i.e. where using the Normal distribution would result in excessive shrinkage towards the mean). 
For example, after \hl{visual inspection of} the variation of frequentist $\emph{orf}\Delta$ level means about their population means in historical datasets, we found many unusually fit, dead or missing $\emph{orf}\Delta$s and concluded that $\emph{orf}\Delta$ fitnesses would be well modelled by the t-distribution.}

Instead of manually fixing the inoculum density parameter $P$ as in \cite{QFA1} our Bayesian hierarchical models deal with the scarcity of information about the early part of culture growth curves by estimating a single $P$ across all $\emph{orf}\Delta$s \hl{(and conditions in some of our models).}
Our new approach learns about $P$ from the data and gives us a posterior distribution to describe our uncertainty about its value.  

\hl{The new, hierarchical structure \citep{BayHi} implemented in our models reflects the structure of QFA experiments.}
Information is shared efficiently among groups of parameters, such as between repeat level parameters for a single mutant strain.
Examples of the type of Bayesian hierarchical modelling which we use to model genetic interaction can be seen in \cite{Zhang2014} and \cite{hierarchical1}, where hierarchical models are used to account for group effects.

In \cite{epis1} the signal of genetic interaction is chosen to be ``strictly ON or OFF" when modelling gene activity. 
\hl{We include this concept in our interaction models by using the posterior probability of a Bernoulli distributed indicator variable \citep{indicator} to describe whether there is evidence of an $\emph{orf}\Delta$ interacting with the query mutation; the more evidence of interaction, the closer posterior expectations will be to one.} 

Failing to account for all sources of variation within the experimental structure, such as the difference in variation between the control and query fitnesses may lead to inaccurate conclusions. 
By incorporating more information into the model with prior distributions and a more flexible modelling approach, we will increase statistical power. 
With an improved analysis it may then be possible for a similar number of genetic interactions to be identified with a smaller sample size (fewer replicate cultures), saving on the experimental costs associated with QFA.

Inference is carried out using Markov Chain Monte Carlo (MCMC) methods. The algorithm used is a Metropolis-within-Gibbs sampler where each full-conditional is sampled in turn either directly or using a simple Normal random walk Metropolis step.  The scheme used is similar to that presented by \cite{Jow2014}.
Due to the large number of model parameters and large quantity of data from high-throughput QFA experiments, the algorithms used \hl{for carrying out inference} often have poor mixing and give highly auto-correlated samples, requiring thinning.
Posterior means are used to obtain point estimates where required.

In the following, we present a two-stage Bayesian, hierarchical modelling approach (Section~\ref{Multiple} and~\ref{sec:interaction}) where we generate $\emph{orf}\Delta$ fitness distributions and infer genetic interaction probabilities separately.  
We then present a one-stage approach (Section~\ref{Joint Model}) for inferring fitness and genetic interaction probabilities simultaneously.  
For the new approaches described in Section \ref{Multiple}, \ref{sec:interaction} and \ref{Joint Model} model fitting is carried out using the techniques discussed above, implemented in C for computational speed, and is freely available in the R package ``qfaBayes'' at \url{https://r-forge.r-project.org/projects/qfa}. 

For the Bayesian models presented, the flow of information within the models and how each parameter is related to the data can be seen from the plate diagrams in Section~1 of the on-line supporting materials.
  \subsection{\label{Multiple}Separate Hierarchical Model}

The separate hierarchical model (SHM), given in (\ref{SHMeqs}), models the growth of multiple yeast cultures using the logistic model described in (\ref{eq:logistic}), whose analytic solution is indicated by $x(t)$.
The observational model at the time point level is given by
\begin{align*}
y_{lmn} &\sim \operatorname{N}(\hat{y}_{lmn},({ \nu_{l}  })^{-1} ) \quad 
&\hat{y}_{lmn} &= x(t_{lmn};K_{lm} ,r_{lm},P),\\
\intertext{where}
l&=1,2,...,L &&\;\;\;\; \text{$\emph{orf}\Delta$ level}\\
m&=1,...,M_{l} &&\;\;\;\; \text{Repeat level}\\
n&=1,2,...,N_{lm} &&\;\;\;\; \text{Time point level.}\\
\intertext{At the next level of the hierarchy (the repeat level), we have}
\log~K_{lm} &\sim \operatorname{N}(K_{l}^o, ({ \tau_{l}^{K} })^{-1} )I_{(-\infty,0]} &\log~\tau_{l}^K &\sim \operatorname{N}(\tau^{K,p}, ({\sigma^{\tau,K}})^{-1} )I_{[0,\infty)}\\
\log~r_{lm} &\sim \operatorname{N}(r_{l}^o, ({ \tau_{l}^{r} })^{-1} )I_{(-\infty,3.5]} &\log~\tau_{l}^r &\sim \operatorname{N}(\tau^{r,p}, ({\sigma^{\tau,r}})^{-1} ).
\end{align*}
Moving up, at the $\emph{orf}\Delta$ level we have
\begin{align*}
e^{K_{l}^o} &\sim t(K^p, ({ \sigma^{K,o} })^{-1},3)I_{[0,\infty)}
&\log~\sigma^{K,o} &\sim \operatorname{N}(\eta^{K,o}, (\psi^{K,o})^{-1} )\\
e^{r_{l}^o} &\sim t(r^p, ({ \sigma^{r,o} })^{-1},3 )I_{[0,\infty)}
&\log~\sigma^{r,o} &\sim \operatorname{N}(\eta^{r,o}, (\psi^{r,o})^{-1} )\\
\log~\nu_{l} &\sim \operatorname{N}(\nu^p, ({ {\sigma}^{\nu} })^{-1} )
&\log~\sigma^{\nu} &\sim \operatorname{N}(\eta^{\nu}, (\psi^{\nu})^{-1} ).\\
\intertext{Finally, at the population level, we take}
\log~K^p &\sim \operatorname{N}(K^\mu, ({\eta^{K,p}})^{-1} )
&\log~r^p &\sim \operatorname{N}(r^\mu, ({\eta^{r,p}})^{-1} )\\
\log~P &\sim \operatorname{N}(P^\mu, ({\eta^{P}})^{-1} )
&\nu^p &\sim \operatorname{N}(\nu^\mu, (\eta^{\nu,p})^{-1} )\\
\tau^{K,p} &\sim \operatorname{N}(\tau^{K,\mu}, ({\eta^{\tau,K,p}})^{-1} )
&\log~\sigma^{\tau,K} &\sim \operatorname{N}(\eta^{\tau,K}, (\psi^{\tau,K})^{-1} )\\
\tau^{r,p} &\sim \operatorname{N}(\tau^{r,\mu}, ({\eta^{\tau,r,p}})^{-1} )
&\log~\sigma^{\tau,r} &\sim \operatorname{N}(\eta^{\tau,r}, (\psi^{\tau,r})^{-1} ).  \numberthis \label{SHMeqs}
\end{align*}

Dependent variable observations $y_{lmn}$ (scaled cell density measurements) and independent variable $t_{lmn}$ (time since inoculation) are model inputs, where $n$ indicates the time point for a given $\emph{orf}\Delta$ repeat.  
A directed, acyclic graph (DAG) for this model can be seen in Section~1 of the supporting on-line information.
In this first hierarchical model, the logistic model is fit to query and control data separately.

In order to measure the variation between $\emph{orf}\Delta$s, parameters ($K^p$,$\sigma^{K}_{o}$) and ($r^p$,$\sigma^{r}_{o}$) are included at the population level of the hierarchy. 
Within-$\emph{orf}\Delta$ variation is modelled by each set of $\emph{orf}\Delta$ level parameters ($K^{o}_{l}$,$\tau^{K}_{l}$) and ($r^{o}_{l}$,$\tau^{r}_{l}$). 
Learning about these higher level parameters allows information to be shared across parameters lower in the hierarchy.
A three-level hierarchical model is applied to $(K,K^{o}_{l},K_{lm})$ and $(r,r^{o}_{l},r_{lm})$, sharing information on the repeat level and the $\emph{orf}\Delta$ level.
Note that $\emph{orf}\Delta$ level parameters $K^{o}_{l}$ and $r^{o}_{l}$ are on the log scale ($e^{K^{o}_{l}}$ and $e^{r^{o}_{l}}$ are on the scale of the observed data).

Assuming a Normal error structure, random measurement error is modelled by the $\nu_l$ parameters (one for each $\emph{orf}\Delta$).
Information on random error is shared across all $\emph{orf}\Delta$s by drawing \hl{$\log \nu_l$} from a Normal distribution parameterised by ($\nu_p$,$\sigma^{\nu}$).
A two-level hierarchical structure is also used for both the $\tau_{l}^{K}$ and $\tau_{l}^{r}$ parameters. 

Modelling logistic model parameter distributions on the log scale ensures that parameters values remain strictly positive (a realistic biological constraint).
Truncating distributions allows us to implement further, realistic constraints on the data. Truncating $\log r_{lm}$ values greater than 3.5 corresponds to disallowing biologically unrealistic culture doubling times \hl{faster than about 30 minutes} and truncating of repeat level parameters $\log K_{lm}$ above 0 ensures that no carrying capacity estimate is greater than the maximum observable cell density, which is 1 after scaling.

\emph{orf}$\Delta$ level parameters $e^{K^o_{l}}$ and $e^{r^{o}_{l}}$ are on the same scale as the observed data.
Realistic biological constraints (positive logistic model parameters) are enforced at the repeat level; however both $e^{K^{l}_{o}}$ and $e^{r_{o}^{l}}$, which are assumed to have scaled~$t$~distributions, are truncated below zero to keep exponentiated parameters strictly positive.

Identifiability problems can arise for parameters $K_{lm}$ and $r_{lm}$ when observed cell densities are low and unchanging (consistent with growth curves for cultures which are very sick, dead or missing).
In these cases, either $K_{lm}$ or $r_{lm}$ can take values near zero, allowing the other parameter to take any value without significantly affecting the model fit. In the \citet{QFA1} approach identification problems are handled in an automated post-processing stage: for cultures with low $K$ estimates (classified as dead), $r$ is automatically set to zero.
Computing time wasted on such identifiability problems is reduced by truncating repeat level parameters $r_{lm}$, preventing the MCMC algorithms from \rev{becoming stuck in} extremely low probability regions when $K_{lm}$ takes near zero values.
Similarly, $\log \tau^{K}_{l}$ parameters are truncated below 0 to overcome identifiability problems between parameters $K_{lm}$ and $r_{lm}$ when $r_{lm}$ takes near zero values.

The SHM (\ref{SHMeqs}) is fit to both the query and control strains separately.  Means are taken to summarise logistic growth parameter posterior distributions. 
Summaries $(\hat{K}_{lm},\hat{r}_{lm},\hat{P})$ for each $\emph{orf}\Delta$ repeat are converted to univariate fitnesses $F_{clm}$ where $c$~identifies the condition (query or control), with any given fitness measure e.g. $D_R\times D_P$ (see (\ref{eq:MDRMDP}) and \cite{QFA1}).  
 
	  \subsection{\label{sec:interaction}Interaction Hierarchical Model}

After the SHM fit, the interaction hierarchical model (IHM), given in (\ref{IHMeqs}) can then be used to model estimated fitness scores $F_{clm}$ and determine, for each $\emph{orf}\Delta$, whether there is evidence for interaction.  

\begin{align*}
c&=0,1  &&\;\;\;\; \text{Condition level}\\
l&=1,...,L_{c} &&\;\;\;\; \text{$\emph{orf}\Delta$ level}\\
m&=1,...,M_{cl} &&\;\;\;\; \text{Repeat level}\\
\intertext{Repeat level}
F_{clm} &\sim \operatorname{N}(\hat{F}_{cl},(\nu_{cl}  )^{-1}) &\hat{F}_{cl} &= e^{\alpha_{c}+Z_{l}+\delta_{l}\gamma_{cl}}\\
\intertext{$\emph{orf}\Delta$ level}
e^{Z_{l}} &\sim t(Z^{p},{({\sigma^{Z}})}^{-1},3)I_{[0,\infty)} &\log~\sigma^{Z} &\sim \operatorname{N}(\eta^{Z},\psi^{Z})\\
\log~\nu_{cl} &\sim \operatorname{N}(\nu^p,{ ({{\sigma}^{\nu}} )}^{-1} ) &\log~\sigma^{\nu} &\sim \operatorname{N}(\eta^{\nu}, \psi^{\nu} )\\
\delta_{l} &\sim Bern(p) &\\
e^{\gamma_{cl}}&=\begin{cases}
1  & \text{if } c=0;\\
t(1,{({\sigma^{\gamma}})}^{-1},3)I_{[0,\infty)} & \text{if } c=1.
\end{cases}
&\log~\sigma^{\gamma}&\sim
\operatorname{N}(\eta^{\gamma},{(\psi^{\gamma})}^{-1})
\\
\intertext{Condition level}
\alpha_{c}&=\begin{cases}
0  & \text{if } c=0;\\
\operatorname{N}(\alpha^\mu,\eta^{\alpha}) & \text{if } c=1.
\end{cases}
\\
\intertext{Population level}
\log~Z^{p} &\sim N(Z^{\mu},{(\eta^{Z,p})}^{-1}) 
&\nu^p &\sim \operatorname{N}(\nu^{\mu}, (\eta^{\nu,p})^{-1} )  \numberthis \label{IHMeqs}
\end{align*}

$F_{clm}$ are the observed fitness scores.
A DAG for this model can be found in Section~1 of the supporting on-line materials.
Fitnesses are passed to the IHM where query screen fitnesses are compared with control screen fitnesses, assuming genetic independence. Deviations from predicted fitnesses are evidence for genetic interaction.  
The interaction model accounts for between $\emph{orf}\Delta$ variation with the set of parameters ($Z^{p}$,$\sigma_{Z}$) and within $\emph{orf}\Delta$ variation by the set of parameters ($Z_{l}$,$\nu_{l}$).
A linear relationship between the control and query $\emph{orf}\Delta$ level parameters is \hl{specified} with a scale parameter $\alpha_{1}$. 
Deviation from this relationship (genetic interaction) is accounted for by the term $\delta_{l}\gamma_{1l}$.
A scaling parameter $\alpha_{1}$ allows any effects due to differences in the control and query data sets to be scaled out, such as differences in \hl{genetic background}, incubator temperature or inoculum density.
The Bernoulli probability parameter $p$ is our prior estimate for the probability of a given $\emph{orf}\Delta$ showing evidence of genetic interaction. For the data set considered in Section~\ref{Application_data} $p$ is set to 0.05 as the experimenter's belief before the experiment was carried out was that $5\%$ of the $\emph{orf}\Delta$s would interact with the query.
Observational noise is quantified by $\nu_{cl}$. 
The $\nu_{cl}$ parameter accounts for difference in variation between condition i.e. the query and control data sets and for difference in variation between $\emph{orf}\Delta$s. 

The linear relationship between the control and query fitness scores, consistent with the multiplicative model of genetic independence, described in (\ref{eq:linear}), is implemented in the IHM as $\hat{F} = e^{\alpha_{c}+Z_{l}+\delta_{l}\gamma_{cl}}=e^{\alpha_{c}}e^{Z_{l}+\delta_{l}\gamma_{cl}}$.
Strains whose fitnesses lie along the linear relationship \hl{defined by} the scalar $\alpha_{1}$ \hl{show no evidence for interaction with the query condition}.  
On the other hand, deviation from the linear relationship, represented by the posterior mean of $\delta_{l}\gamma_{1l}$ \hl{is} evidence for genetic interaction.
The larger the posterior mean for $\delta_{l}$ is, the \hl{higher} the probability or evidence there is for interaction, \hl{while $\gamma_{1l}$ is a measure of the strength of interaction}. 
Where the query condition has a negative effect (i.e. decreases fitness on average, compared to the control condition), query fitnesses which are above and below the linear relationship are suppressors and enhancers of the fitness defect associated with the query condition respectively.
A list of genes ranked by strength and direction of interaction with the query condition are ordered by the posterior means of $\delta_{l}\gamma_{cl}$.
The $\emph{orf}\Delta$s with $\hat{\delta}_{l}>0.5$ are classified and labelled as showing ``significant'' evidence of interaction.
 
 \subsection{\label{Joint Model}Joint Hierarchical Model}

The joint hierarchical model (JHM), given in (\ref{JHMeqs}) is an alternative, fully Bayesian version of the two-stage approach described in Section \ref{Multiple} and \ref{sec:interaction}.

\begin{align*}
c&=0,1   &&\;\;\;\; \text{Condition level}\\
l&=1,...,L_{c}   &&\;\;\;\; \text{$\emph{orf}\Delta$ level}\\
m&=1,...,M_{cl}       &&\;\;\;\; \text{Repeat level}\\
n&=1,...,N_{clm}      &&\;\;\;\; \text{Time point level}\\
\intertext{Time point level}
y_{clmn} &\sim \operatorname{N}(\hat{y}_{clmn},({\nu_{cl}})^{-1} )\quad
&\hat{y}_{clmn} &= x(t_{clmn};{ K_{clm} } ,{ r_{clm} } , { P })\\
\intertext{Repeat level}
\log~K_{clm} &\sim \operatorname{N}(\alpha_{c}+K_{l}^o+\delta_{l}\gamma_{cl},({ \tau_{cl}^K })^{-1})I_{(-\infty,0]}
\qquad & \log~\tau_{cl}^K &\sim \operatorname{N}(\tau^{K,p}_{c}, ({\sigma^{\tau,K}_{c}})^{-1} )I_{[0,\infty)}\\
\log~r_{clm} &\sim \operatorname{N}(\beta_{c}+r_{l}^o+\delta_{l}\omega_{cl},({ \tau_{cl}^r })^{-1})I_{(-\infty,3.5]}
\qquad &\log~\tau_{cl}^r &\sim \operatorname{N}(\tau^{r,p}_{c}, ({\sigma^{\tau,r}_{c}})^{-1} )\\
\intertext{$\emph{orf}\Delta$ level}
e^{K_{l}^o} &\sim t(K^p, ({ \sigma^{K,o} })^{-1},3 )I_{[0,\infty)}\qquad &\log~\sigma^{K,o} &\sim \operatorname{N}(\eta^{K,o}, (\psi^{K,o})^{-1} )\\
e^{r_{l}^o} &\sim t(r^p, ({ \sigma^{r,o} })^{-1},3 )I_{[0,\infty)}\qquad &\log~\sigma^{r,o} &\sim \operatorname{N}(\eta^{r,o}, (\psi^{r,o})^{-1} )\\
\log~\nu_{cl} &\sim \operatorname{N}(\nu^p,({ \sigma^{\nu} })^{-1})\qquad& \log~\sigma^{\nu} &\sim \operatorname{N}(\eta^{\nu}, (\psi^{\nu})^{-1} )\\
\delta_{l} &\sim Bern(p)\\
e^{\gamma_{cl}}&=\begin{cases}
1  & \text{if } c=0;\\
t(1,{({\sigma^{\gamma}})}^{-1},3)I_{[0,\infty)} & \text{if } c=1.
\end{cases}
\qquad
&\log~\sigma^{\gamma}&\sim
\operatorname{N}(\eta^{\gamma},\psi^{\gamma})  
\\
e^{\omega_{cl}}&=\begin{cases}
1  & \text{if } c=0;\\
t(1,({{\sigma^{\omega}})}^{-1},3)I_{[0,\infty)} & \text{if } c=1.
\end{cases}
\qquad
&\log~\sigma^{\omega}&\sim
\operatorname{N}(\eta^{\omega},\psi^{\omega})
\\
\intertext{Condition level}
\alpha_{c}&=\begin{cases}
0  & \text{if } c=0;\\
\operatorname{N}(\alpha^{\mu},\eta^{\alpha})  & \text{if } c=1.
\end{cases}
&\beta_{c}&=\begin{cases}
0  & \text{if } c=0;\\
\operatorname{N}(\beta^{\mu},\eta^{\beta}) & \text{if } c=1.
\end{cases}
\\
\tau^{K,p}_{c} &\sim \operatorname{N}(\tau^{K,\mu}, ({\eta^{\tau,K,p}})^{-1} )
& \log~\sigma^{\tau,K}_{c} &\sim \operatorname{N}(\eta^{\tau,K}, (\psi^{\tau,K})^{-1} )\\
\tau^{r,p}_{c} &\sim \operatorname{N}(\tau^{r,\mu}, ({\eta^{\tau,r,p}})^{-1} )
& \log~\sigma^{\tau,r}_{c} &\sim \operatorname{N}(\eta^{\tau,r}, (\psi^{\tau,r})^{-1} )\\
\intertext{Population level}
\log~K^p &\sim \operatorname{N}(K^\mu, ({\eta^{K,p}})^{-1} )
& \log~r^p &\sim \operatorname{N}(r^\mu, ({\eta^{r,p}})^{-1} )\\
\nu^p &\sim \operatorname{N}(\nu^\mu, ({\eta}^{\nu,p})^{-1} )
& \log~P &\sim \operatorname{N}(P^\mu, ({\eta^{P}})^{-1} ) \numberthis \label{JHMeqs}
\end{align*}

Here, the dependent variable $y_{clmn}$ (scaled cell density measurements) and independent variable $t_{clmn}$ (time since inoculation) are input to the JHM.
The JHM incorporates the key modelling ideas from both the SHM and the IHM with the considerable advantage that we can learn about logistic growth model, fitness and genetic interaction parameters simultaneously, thereby avoiding having to choose a fitness measure or point estimates for passing information between models. 
The JHM is an extension of the SHM with the presence or absence of genetic interaction being described by a Bernoulli indicator and an additional level of error to account for variation due to the query condition.  
Genetic interaction is modelled in terms of the two logistic growth parameters $K$ and $r$ simultaneously.

By fitting a single JHM, we need only calculate posterior means, check model diagnostics and thin posterior samples once. However, the CPU time taken to reach convergence for any given data set is roughly twice that of the two-stage approach for a genome-wide QFA.

All of the SHM and IHM modelling assumptions described in Sections~\ref{Multiple}~and~\ref{sec:interaction}, such as distributional choices and hierarchical structure are inherited by the JHM.
Similar to the interaction model in Section~\ref{sec:interaction}, linear relationships between control and query carrying capacity and growth rate (instead of fitness score) are assumed:  $(e^{\alpha_{c}+K^{o}_{l}+\delta_{l}\gamma_{cl}},e^{\beta_{c}+r^{o}_{l}+\delta_{l}\omega_{cl}})$.

\section{\label{Application_data}Re-analysis of QFA experiments designed to learn about telomere biology}

In this section we present a re-analysis of a previously published experiment, designed to inform us about the ways that eukaryotic cells respond to the loss of telomere caps that normally protect the ends of chromosomes from being erroneously recognised as a type of DNA damage.
A pair of genome-wide QFA screens, were carried out in the model eukaryotic organism \emph{S. cerevisiae} (brewer's yeast), comparing the fitness of control \emph{ura3$\Delta$} strains with query \mbox{\emph{cdc13-1}} \emph strains.
These comparisons were made to identify genes that show evidence of interaction with the query mutation \mbox{\emph{cdc13-1}}. 
Cdc13 is a \emph{S. cerevisiae} protein which binds to telomeres and regulates telomere capping.
\mbox{\emph{cdc13-1}} is a temperature-sensitive allele of the \emph{CDC13} gene.
The ability of the altered Cdc13 protein produced by strains carrying the \mbox{\emph{cdc13-1}} gene to cap telomeres is reduced at temperatures above $26\,^{\circ}\mathrm{C}$ \citep{cdc131}, inducing a fitness defect that can be measured by QFA. 
The original experimental data used are freely available at \url{http://research.ncl.ac.uk/colonyzer/Addi​nallQFA/}.
\cite{QFA1} present a list of inferred interaction strengths and p-values for significance of interaction, together with a fitness plot for this experiment. 
 
Here, we will compare lists of genes classified as interacting with \mbox{\emph{cdc13-1}} by the non-hierarchical frequentist approach presented by \cite{QFA1} and the hierarchical REM with those classified as interacting by our hierarchical Bayesian approaches. 

4,294 non-essential $\emph{orf}\Delta$s were selected from the yeast deletion collection and used to build the corresponding double deletion control and query strains. 
Independent replicate culture growth curves (time course observations of cell density) were captured for each control and query strain. 
The median and range for the number of replicates per $\emph{orf}\Delta$ is 8 and $[8,144]$ respectively.
The range for the number of time points for growth curves captured in the control experiment is $[7,22]$ and $[9,15]$ in the query experiment.  

As in the \cite{QFA1} analysis, a list of 159 genes are stripped from our final list of genes for biological and experimental reasons. 
Priors for the models used throughout Section \ref{Application_data} are provided in Table \ref{table:priors}.
We have ensured that these priors are sufficiently diffuse to describe any QFA data set by inspecting 10 historical QFA data sets.

\begin{table}
\caption{\label{table:priors}Hyperparameter values specifying priors for the SHM, IHM and JHM.}
\resizebox{!}{0.9in}{%
\npdecimalsign{.}
\nprounddigits{2}
\centering 
\begin{tabular}{c n{2}{2} c n{2}{2} c n{2}{2} c n{2}{2}}
    \hline
\noalign{\vskip 0.4mm} 
		\multicolumn{2}{c}{SHM \& JHM} &\multicolumn{2}{c}{SHM \& JHM}& \multicolumn{2}{c}{JHM} & \multicolumn{2}{c}{IHM}\\
Parameter Name  & \multicolumn{1}{c}{Value} & Parameter Name  & \multicolumn{1}{c}{Value} & Parameter Name  & \multicolumn{1}{c}{Value}& Parameter Name  & \multicolumn{1}{c}{Value}\\ \hline
$\tau^{K,\mu}$ & 2.20064039227566   & $\eta^{r,p}$ & 0.133208648543871 &$\alpha^{\mu}$ & 0 & $Z_{\mu}$ & 3.65544229414228  \\ 
$\eta^{\tau,K,p}$ & 0.0239817523340161  &       $\nu^{\mu}$ & 19.8220570630669 &  $\eta^{\alpha}$ & 0.25 & $\eta^{Z,p}$ & 0.697331530063874 \\  
 $\eta^{K,o}$ & -0.79421175992029 & $\eta^{\nu,p}$ & 0.0174869367984725 &$\beta^{\mu}$ & 0 &  $\eta^{Z}$ & 0.104929506383255 \\ 
  $\psi^{K,o}$ & 0.610871036009521 &  $P^{\mu}$ & -9.03928728018792 &  $\eta^{\beta}$ & 0.25  &   $\psi^{Z}$ & 0.417096744759774 \\  
     $\tau^{r,\mu}$ &  3.64993037268256 &  $\eta^{P}$ & 0.469209463148874 &  $p$ & 0.05 & $\eta^{\nu}$ & 0.101545024587153 \\   
 $\eta^{\tau,r,p}$ &  0.0188443648965434 &&&$\eta^{\gamma}$ & -0.79421175992029 &  $\psi^{\nu}$ & 2.45077729037385 \\ 
$\eta^{r,o}$ & 0.468382435659566  &&& $\psi^{\gamma}$ & 0.610871036009521 &    $\nu^{\mu}$ & 2.60267545154548 \\ 
 $\psi^{r,o}$ & 0.0985295312016232  &&&  $\eta^{\omega}$ & 0.468382435659566 &    $\eta^{\nu,p}$ &0.0503202367841729  \\ 
  $\eta^{\nu}$ & -0.834166609695065  &&& $\psi^{\omega}$ & 0.0985295312016232 &    $\alpha^{\mu}$ & 0 \\ 
 $\psi^{\nu}$ & 0.855886535578262  &&& $\eta^{\tau,K}$ &  2.20064039227566&    $\eta^{\alpha}$ & 0.309096075088720 \\   
  $K^{\mu}$ & -2.01259579112252  &&&   $\psi^{\tau,K}$ & 0.0239817523340161 &    $p$ & 0.05 \\ 
 $\eta^{K,p}$ & 0.032182397822033  &&& $\eta^{\tau,r}$ & 3.64993037268256 & $\eta^{\gamma}$ & 0.104929506383255 \\ 
    $r^{\mu}$ & 0.97398228941848  &&&   $\psi^{\tau,r}$ & 0.0188443648965434 &  $\psi^{\gamma}$ & 0.417096744759774 \\  
    \hline
    \end{tabular}
    \npnoround
		}
\end{table}

\subsection{\label{Application}Model application}

\hl{The Heidelberg-Welch \citep{Heidelberger} and Raftery-Lewis \citep{Raftery} convergence diagnostics are used to determine whether convergence has been reached for all parameters.
Posterior and prior densities are compared by eye to ensure that sample posterior distributions are not restricted by the choice of prior distribution.
ACF (auto-correlation) plot diagnostics are checked visually to ensure that serial correlation between sample values of the posterior distribution is low, ensuring that the effective sample size is similar to the actual sample size.} 

\begin{figure}[h!]
  \centering
\includegraphics[width=14cm]{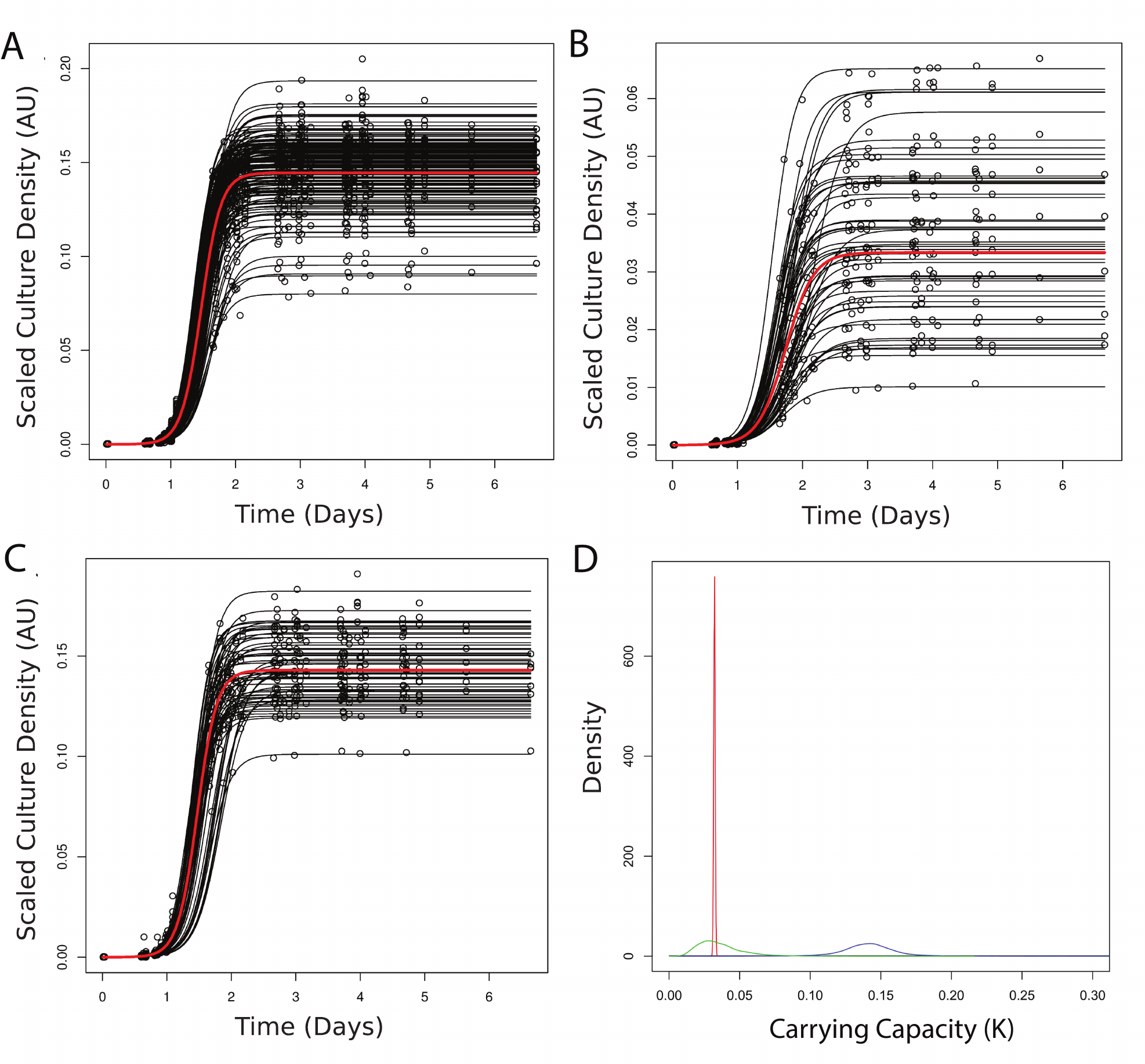}
 \caption{
Hierarchy of model fits and parameter estimates.
Data for $\emph{orf}\Delta$ repeats have been plotted in A, B and C, with SHM fitted curves overlaid in black for repeat level parameters and red for the $\emph{orf}\Delta$ level parameter fit. 
A) SHM scatter plot for 144 \emph{his3}$\Delta$ \emph{ura3$\Delta$} repeats at~${27}^{\circ}$C. 
B) SHM scatter plot for 48 \emph{rad50}$\Delta$ \emph{ura3$\Delta$} repeats at~${27}^{\circ}$C. 
C) SHM scatter plot for 56 \emph{exo1}$\Delta$ \emph{ura3$\Delta$} repeats at~${27}^{\circ}$C. 
D) SHM density plot of posterior predictive distributions for \emph{rad50}$\Delta$ \emph{ura3$\Delta$} 
carrying capacity $K$ hierarchy. 
The prior distribution for $K^p$ is flat over this range.
The posterior predictive for $e^{K^o_l}$ is in blue and for $K_{clm}$ in green.
The posterior distribution of the first time course repeat $K_{clm}$ parameter is in red.
Parameters $K^p$, $e^{K^o_l}$ and $K_{clm}$ are on the same scale as the observed data.
}

\label{fig:diagABC}
\end{figure}

To assess how well the logistic growth model describes cell density observations we generate plots of raw data with fitted curves overlaid. 
Figure~\ref{fig:diagABC}A, \ref{fig:diagABC}B and \ref{fig:diagABC}C show time series data for three different mutant strain repeats at~${27}^{\circ}$C, together with fitted logistic curves.
Alternative fitness plots can be found in Section 3 of the online supporting material.
We can see that each $\emph{orf}\Delta$ curve fit represents the repeat level estimates well as each $\emph{orf}\Delta$ level (red) curve lies in the region where most repeat level (black) curves are found.
Sharing information between $\emph{orf}\Delta$s will also affect each $\emph{orf}\Delta$ curve fit, increasing the probability of the $\emph{orf}\Delta$ level parameters being closer to the population parameters.
Comparing Figure~\ref{fig:diagABC}A, \ref{fig:diagABC}B and \ref{fig:diagABC}C shows that the SHM \hl{captures heterogeneity at} both the repeat and $\emph{orf}\Delta$ levels.

Figure~\ref{fig:diagABC}D demonstrates the hierarchy of information about \hl{the logistic model parameter $K$} generated by the \hl{SHM} for the $\emph{rad50}\Delta$ control mutant strain (variation decreases going from population level down to repeat level). 
Figure~\ref{fig:diagABC}D \hl{also shows that the posterior distribution for $K$ is much more peaked than the prior, demonstrating that we have learned about the distribution of both the population and $\emph{orf}\Delta$ parameters.}
Learning more about the repeat level parameters \hl{reduces} the variance of our $\emph{orf}\Delta$ level estimates.
The posterior for the first time course repeat \hl{ $K_{clm}$ } parameter shows exactly how much uncertainty there is for this particular repeat in terms of carrying capacity $K$.

\subsubsection{\label{sec:epiplot}Fitness Plots}

Fitness plots are used to show which $\emph{orf}\Delta$s show evidence of genetic interaction. 
The plots are typically mean $\emph{orf}\Delta$ fitnesses for query strains against the corresponding control strains.

\begin{figure}[h!]
  \centering
\includegraphics[width=14cm]{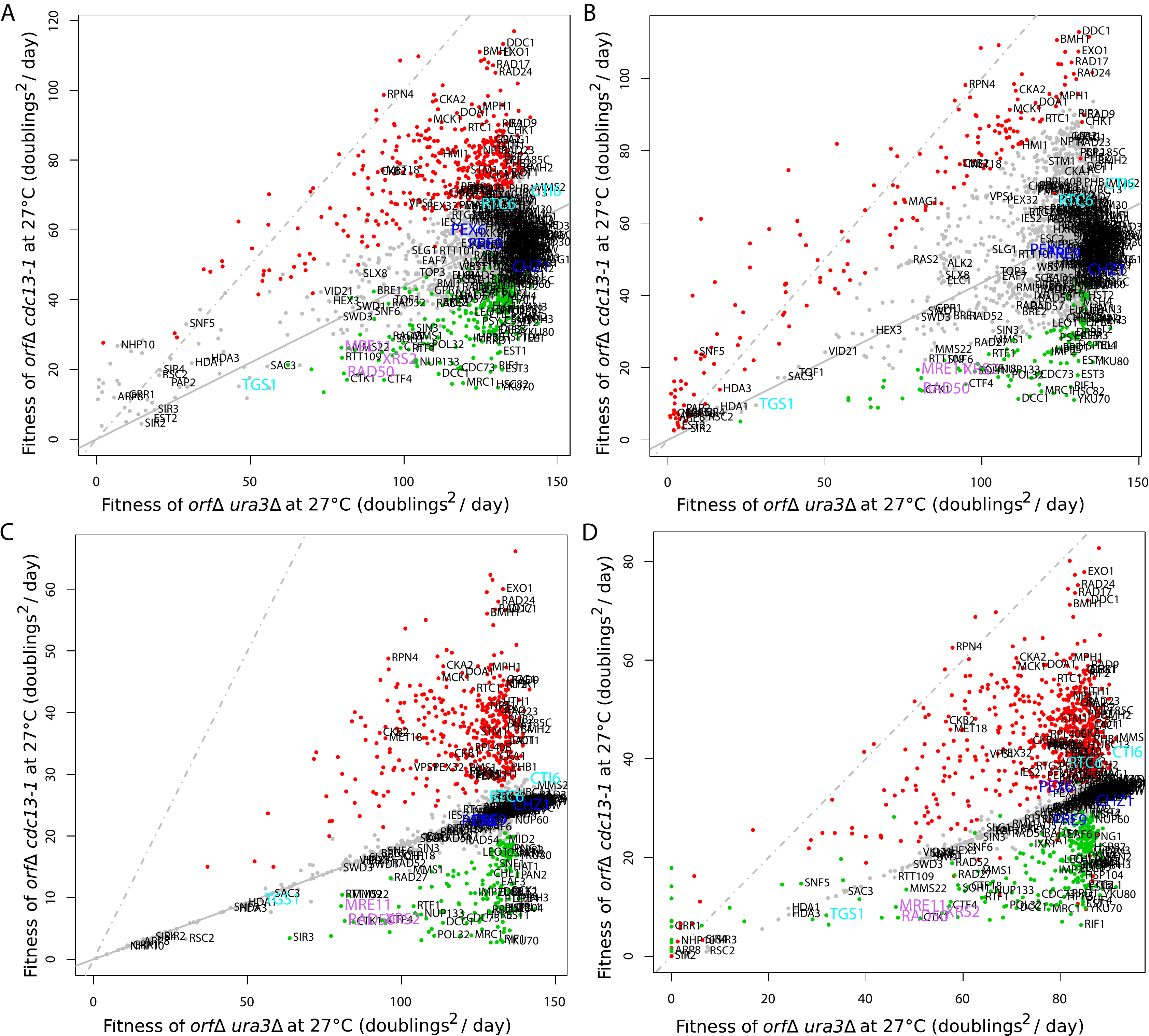}
\caption{Fitness plots comparing mean fitnesses $(F=D_R\times D_P)$ for each \emph{orf}$\Delta$ in a query and control screen.
$\emph{orf}\Delta$s significantly suppressing or enhancing the \emph{cdc13-1} fitness defect are highlighted in red and green respectively.
A) Non-Bayesian, non-hierarchical fitness plot, based on Table~S6 from \cite{QFA1}.
B) Non-Bayesian, hierarchical fitness plot, from fitting REM to data in Table~S6 from \cite{QFA1}.
C) IHM fitness plot. 
D) JHM fitness plot, $\emph{orf}\Delta$s are classified as suppressors or enhancers based on analysis of growth parameter $r$: some strains are fitter in the query experiment than predicted based on control, but are classified as enhancers (green).
A \& B: significant interactors are classified as those with FDR corrected p-values $<0.05$.
C \& D: significant interactors have posterior probability $\delta_l>0.5$.
Labelled genes are annotated with GO terms from Table~\ref{tab:sup_enh}: ``telomere maintenance'', ``ageing'', ``response to DNA damage stimulus'' or ``peroxisomal organization'', as well as genes identified as interactions using the JHM by considering $K$ (see Figure~\ref{fig:JHM_only}) (blue) or by considering $r$ (cyan) and the MRX complex genes (pink).
Solid and dashed grey fitted lines are the line of equal fitness and linear model fits respectively.
\vspace{0.2in}
}
\label{fig:old}
\label{fig:REM} 
\label{fig:IHM}
\label{fig:JHM}
\end{figure}

Figure~\ref{fig:old}A is a fitness plot from \cite{QFA1} where growth curves and evidence for genetic interaction are modelled using the frequentist, non-hierarchical methodology discussed in Section~\ref{sec:previous}.
Figure~\ref{fig:REM}B is a fitness plot for the frequentist hierarchical approach REM, described in \ref{REMeqs}, applied to the logistic growth parameter estimates used in \cite{QFA1}.
The number of genes identified as interacting with \emph{cdc13-1} by \cite{QFA1} and by the REM are \rev{715 and 315} respectively (Table~\ref{tab:sup_enh}).
The REM has highlighted many strains which have low fitness.
In order to fit a linear model to the fitness data and interpret results in terms of the multiplicative model we apply a log transformation to the fitnesses, thereby affecting the distribution of $\emph{orf}\Delta$ level variation.

The REM accounts for between subject variation and allows for the estimation of a query mutation and $\emph{orf}\Delta$ effect to be made simultaneously, unlike the model presented by \cite{QFA1}.
Due to the limitations of the frequentist \hl{hierarchical} modelling framework, the REM model assumes equal variances for all $orf\Delta$s and incorrectly describes \emph{orf}$\Delta$ level variation as Log-normal, assumptions that are not necessary in our new Bayesian approaches.

\clearpage

\begin{table}
\caption{\label{tab:sup_enh}Number of genes interacting with \emph{cdc13-1} at ${27}^{\circ}$C identified using each of four approaches: Add \citep{QFA1}, REM, IHM and JHM.
 Number of genes classified (or annotated) with four example GO terms (telomere maintenance, ageing, response to DNA damage stimulus and peroxisome organisation) are also listed.
For the Add and REM approach, significant interactors are classified as those with FDR corrected p-values (q-values) $<0.05$.
The label ``half data'' denotes analyses where only half of the available experimental observations are used.
}
\centering
\includegraphics[width=14cm]{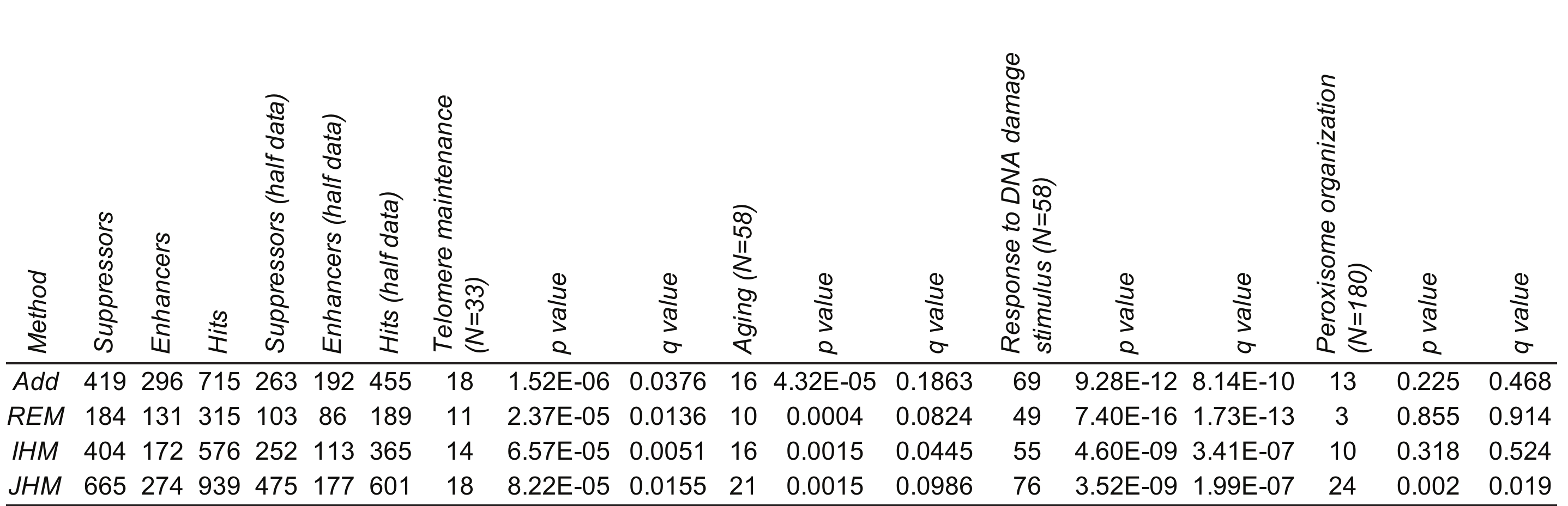}
\end{table}

\subsection{\label{Application1}Application of the two-stage modelling procedure to a suppressor/enhancer data set}

Figure \ref{fig:IHM}C is an IHM fitness plot with $\emph{orf}\Delta$ level fitness measures generated using the new Bayesian two-stage methodology with fitness in terms of $D_R\times D_P$. 
576 genes are identified by the IHM as genetic interactions (Table~\ref{tab:sup_enh}).
Logistic parameter posterior means are used to generate fitness measures. For a gene $(l)$ from the gene deletion library, $(e^{Z_{l}})$ is the fitness for the control and $(e^{\alpha_{1}+Z_{l}+\delta_{l}\gamma_{c,l}})$ for the query.
Similar to Figure~\ref{fig:old}A~and~\ref{fig:REM}B, Figure \ref{fig:IHM}C shows how the majority of control strains are more fit than their query strain counterparts, with a mean fitted line lying below the line of equal fitness. 
Comparing the fitted lines in Figure~\ref{fig:old}A~and~\ref{fig:REM}B with Figure~\ref{fig:IHM}C, the IHM shows the largest deviation between the fitted line and the line of equal fitness, is largely due to the difference in $P$ estimated with the SHM for the control and query data sets being scaled out by the parameter $\alpha_{1}$.
 \rev{
If we fix $P$ in our Bayesian models, as in the frequentist approach, genetic interactions identified are similar, but we then have the problem of choosing $P$. 
We recommend estimating $P$ simultaneously with the other model parameters because if the choice of $P$ is not close to the true value, growth rate $r$ estimates must compensate and do not give accurate estimates for time courses with low carrying capacity $K$.
}

It can be seen that many of the interacting $\emph{orf}\Delta$s have large deviations from the genetic independence line. 
This is because of the indicator variable in the model, used to describe genetic interaction. 
When there is enough evidence for interaction the binary variable is set to 1, otherwise it is set to 0. 
It is interesting to note that non-significant $\emph{orf}\Delta$s, marked by grey points, lie amongst some of the significant strains. 
Many \hl{such points} have high variance and we are therefore \hl{less confident that these interact with the query mutation}.
This \hl{feature} of our new approach \hl{is an improvement over that} presented in \cite{QFA1}, which always shows evidence for an epistatic effect, for a given number of replicates, when mean distance from the genetic independence line is large, regardless of actual strain fitness variability.

\subsection{\label{Application2}Application of the Joint \hl{Hierarchical Model} to a suppressor/enhancer data set }
Figure~\ref{fig:JHM}D is a JHM $D_R\times D_P$ fitness plot using the new, \hl{unified} Bayesian methodology.
939 genes are identified by the JHM as genetic interactions (Table~\ref{tab:sup_enh}).
Posterior means of model parameters are used to obtain the following fitness measures. 
For a gene ($l$) from the gene deletion library, $(e^{K^{o}_{l}},e^{r^{o}_{l}})$ are used to evaluate the fitness for the control and $(e^{\alpha_{1}+K^{o}_{l}+\delta_{l}\gamma_{c,l}},e^{\beta_{1}+r^{o}_{l}+\delta_{l}\omega_{c,l}})$ for the query.
 

Instead of producing a fitness plot in terms of $D_R\times D_P$, it can also be useful to analyse carrying capacity $K$ and growth rate $r$ fitness plots as, in the JHM, evidence for genetic interaction comes from both of these parameters simultaneously.
Fitness plots in terms of logistic growth parameters are useful for identifying some unusual characteristics of $\emph{orf}\Delta$s.
For example, an $\emph{orf}\Delta$ may be defined as a suppressor in terms of $K$ but an enhancer in terms of $r$.
To enable direct comparison with the \cite{QFA1} analyses we generated a $D_R\times D_P$ fitness plot, Figure~\ref{fig:JHM}D.

\subsection{\label{Application3}Comparison with previous analysis}

\subsubsection{\label{individual_interactions}Significant genetic interactions}
\rev{
Of the genes identified as interacting with \emph{cdc13-1} some are identified consistently across all four approaches (215 out of 1038, see Table~\ref{tab:overlap}A).  Of the hits identified by the JHM (939), the majority (639) are common with those in the previously published \citet{QFA1} approach.  However, 231 of 939 are uniquely identified by the JHM and could be interesting candidates for further study.}

\begin{figure}[h!]
  \centering
\includegraphics[width=14cm]{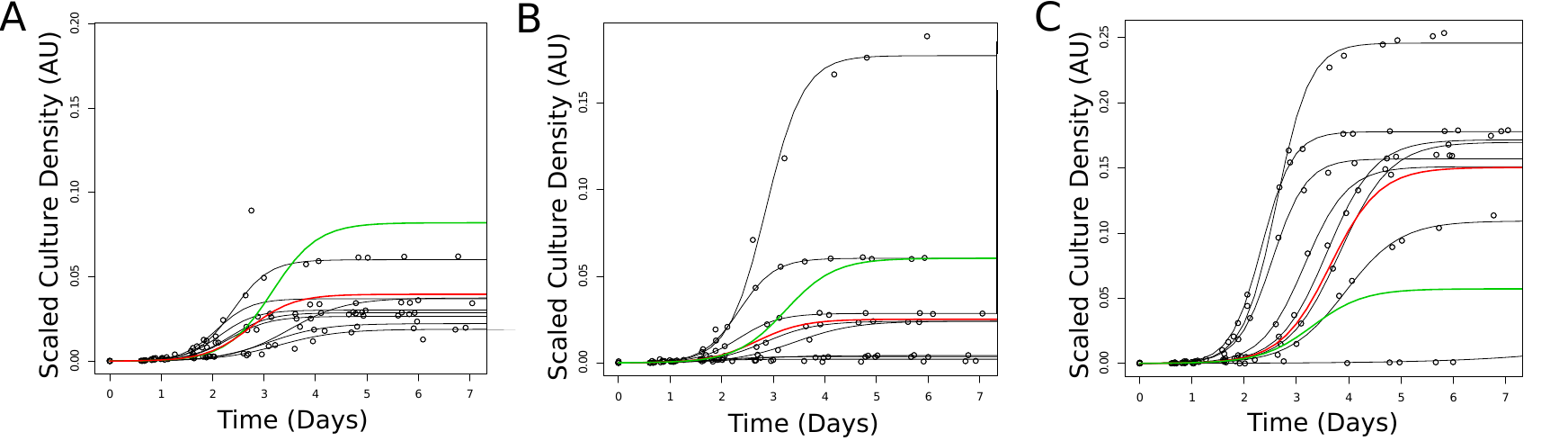}
 \caption{
Hierarchy of growth curve model fits for the JHM for some example genotypes.
JHM data for $\emph{orf}\Delta$ repeats have been plotted in A, B and C, with fitted curves overlaid in black for repeat level parameters, red for the  $\emph{orf}\Delta$ level query parameter fit and green for the expected $\emph{orf}\Delta$ level query parameter fit with no genetic interaction. 
A) JHM scatter plot for 8 \emph{chz1}$\Delta$ \emph{cdc13-1} repeats. 
B) JHM scatter plot for 8 \emph{pre9}$\Delta$ \emph{cdc13-1} repeats. 
C) JHM scatter plot for 8 \emph{pex6}$\Delta$ \emph{cdc13-1} repeats. 
}
\label{fig:JHM_only}
\end{figure}

To examine the evidence for some interactions uniquely identified by the JHM in more detail we compared the growth curves for three examples from the group of interactions identified only by the JHM.  These examples (\emph{chz1}$\Delta$, \emph{pre9}$\Delta$ and \emph{pex6}$\Delta$) are genetic interactions which can be identified in terms of carrying capacity $K$, but not in terms of growth rate $r$ (a unique feature of the JHM, see Figure~\ref{fig:JHM_only}). 
By observing the difference between the fitted growth curve (red) and the expected growth curve, given no interaction (green) in Figure~\ref{fig:JHM_only}A, \ref{fig:JHM_only}B and \ref{fig:JHM_only}C we test for genetic interaction.  Since the expected growth curves in the absence of genetic interaction are not representative of either the data or the fitted curves on the repeat and \emph{orf}$\Delta$ level, there is evidence for genetic interaction.

\begin{table}
\caption{\label{tab:overlap}Genes interacting with \emph{cdc13-1} and GO terms over-represented in list of interactions according to each approach A) Number of genes identified for each approach (Add, REM, IHM and JHM) and the overlap between the approaches. 4135 genes from the \emph{S. cerevisiae} single deletion library are considered.
B) Number of GO terms identified for each approach and the overlap between the approaches.  6107 \emph{S. cerevisiae} GO Terms available.}
\centering
\resizebox{\columnwidth}{!}{%
\begin{tabular}{*{6}{c}}
\multicolumn{1}{l}{\bf{A.}}&  & \multicolumn{2}{c}{\emph{REM:0}} & \multicolumn{2}{c}{\emph{REM:1}}\\
\cline{3-6}
& &\emph{Add:0} &\emph{Add:1} &\emph{Add:0} &\emph{Add:1}\\\hline
\multirow{2}{*}{\emph{IHM:0}} &\emph{JHM:0} &3097&54&31&10\\
& \emph{JHM:1} &231&78&29&29\\\hline
\multirow{2}{*}{\emph{IHM:1}} &\emph{JHM:0} &1&2&1&0\\
&\emph{JHM:1} &30&327&0&215\\\hline
\end{tabular}
\qquad
\begin{tabular}{*{6}{c}}
\multicolumn{1}{l}{\bf{B.}}&  & \multicolumn{2}{c}{\emph{REM:0}} & \multicolumn{2}{c}{\emph{REM:1}}\\
\cline{3-6}
& &\emph{Add:0} &\emph{Add:1} &\emph{Add:0} &\emph{Add:1}\\\hline
\multirow{2}{*}{\emph{IHM:0}} &\emph{JHM:0} &5813&21&58&7\\
& \emph{JHM:1} &46&8&6&10\\\hline
\multirow{2}{*}{\emph{IHM:1}} &\emph{JHM:0} &20&15&3&12\\
&\emph{JHM:1} &13&54&2&147\\\hline
\end{tabular}
}
\end{table}

We chose a prior for the probability $p$ of a gene interacting with the background mutation as 0.05, and explore the effect of alternate choices below.
We therefore expected to find 215 genes interacting.
Using the Bayesian models, for which a prior is applicable (IHM and JHM), we find more genes than expected (576 and 939 interactions respectively, Table~\ref{tab:sup_enh}), demonstrating that this dataset is sufficiently information rich to overcome prior expectations.
The JHM identifies the highest proportion of genes as hits out of all methods considered, particularly identifying suppressors of \emph{cdc13-1} (Table~\ref{tab:sup_enh}).
In fact, the JHM identifies more hits than the \citet{QFA1} approach, even when constrained to using only half of the available data.
An important advantage of our new Bayesian approaches is that we no longer have to choose a q-value threshold.
For the \citet{QFA1} approach to have similar numbers of interactions to the JHM, a less stringent q-value threshold would have to be justified \emph{a posteriori} by the experimenter.

\subsubsection{\label{known_interactions}Previously known genetic interactions}
In order to compare the quality of our new, Bayesian hierarchical models with existing, frequentist alternatives, we examined the lists of genetic interactions identified by all the methods discussed and presented here.
Comparing results with expected or \hl{previously} known lists of interactions from the relevant literature, we find that genes coding for the MRX complex (\emph{MRE11}, \emph{XRS2} \& \emph{RAD50}), which are known to interact subtly with \emph{cdc13-1} \citep{MRX}, are identified by all four approaches considered and can be seen in a similar position on all four fitness plots (Figure~\ref{fig:old}A, \ref{fig:REM}B, \ref{fig:IHM}C and \ref{fig:JHM}D).

By observing the genes labelled in Figure~\ref{fig:old}A~and~\ref{fig:REM}B we can see that the frequentist approaches are unable to identify many of the interesting genes identified by the JHM as these methods are unable to detect interactions for genes close to the genetic independence line.
It seems likely that the JHM has extracted more information from deletion strain fitnesses observed with high variability than the \cite{QFA1} approach by sharing more information between levels of the hierarchy, consequently improving our ability to identify interactions for genes that are found closer to the line of genetic independence (subtle interactions).  \emph{CTI6}, \emph{RTC6} and  \emph{TGS1} are three examples of subtle interactors identified only by the JHM (interaction in terms of $r$ but not $K$) which all have previously known telomere-related functions \citep{TGS1,CTI6,RTC6}.

We tested the biological relevance of results from the various approaches by carrying out unbiased Gene Ontology (GO) term enrichment analyses on the hits (lists of genes classified as having a significant interaction with \emph{cdc13-1}) using the \rev{bioconductoR package GOstats \citep{GOstats}} (see Section~2 of the on-line supporting materials).
As an example, fitness plots with genes co-annotated with the ``telomere maintenance'' highlighted can be seen in Section~3 of the on-line supporting materials. 

\rev{Extracts from the list of top interactions identified by both the IHM and JHM are provided in Section~4 of the on-line supporting materials.
Files including the full lists of genetic interactions for the IHM and JHM are also provided (\url{http://research.ncl.ac.uk/qfa/HeydariQFABayes/}).}
Since we can use the JHM to identify interactions in terms of both $K$ and $r$ simultaneously, it is useful to order lists of suppressors and enhancers in terms of $K$ and $r$ as well as a fitness measure such as $D_R\times D_P$ for reviewing the results, see Section~5 of the on-line supporting materials. 

All methods identify a large proportion of the genes in the yeast genome annotated with the GO terms ``telomere maintenance'' and ``response to DNA damage stimulus'' (see Table~\ref{tab:sup_enh} and the on-line supporting materials.), which were the targets of the original screen, demonstrating that they all correctly identify previously known hits of biological relevance.  Interestingly, the JHM identifies many more genes annotated with the ``ageing'' GO term, which we also expect to be related to telomere biology (though the role of telomeres in ageing remains controversial) suggesting that the JHM is identifying novel, relevant interactions not previously identified by the \citet{QFA1} screen (see Table~\ref{tab:sup_enh}).  Similarly, the JHM identifies a much larger proportion of the PEX ``peroxisomal'' complex (included in GO term: ``peroxisome organisation'') as interacting with \emph{cdc13-1} (see Table~\ref{tab:sup_enh}) including all of those identified in \citet{QFA1}. Many of the PEX genes show large variation in both $K$ and $r$, an example can be seen in Figure~\ref{fig:JHM_only}C for \emph{pex6$\Delta$}. Members of the PEX complex cluster tightly, above the fitted line in the fitness plot Figure~\ref{fig:JHM}D \rev{(fitness plots with highlighted genes for GO terms in Table~\ref{tab:sup_enh} are given in Section~3 of the on-line supporting materials)}, demonstrating that although these functionally related genes are not strong interactors, the same behaviour is reproduced independently by multiple members of a known functional complex, suggesting that the predicted interactions are real.  The results of tests for significant over-representation of all GO terms can be found online: \url{http://research.ncl.ac.uk/qfa/HeydariQFABayes/}.

Overall, within the lists of genes identified as interacting with \emph{cdc13-1} by the \citet{QFA1}, REM, IHM and JHM approaches, 274, 245, 266 and 286 GO terms were significantly over-represented respectively (out of 6235 possible GO terms, see Table \ref{tab:overlap}B).  147 were common to all approaches and examples from the group of GO terms over-represented in the JHM analysis and not in the \citet{QFA1} analysis seem internally consistent (e.g. ``peroxisome organisation'' GO term) and consistent with the biological target of the screen, telomere biology (significant GO terms for genes identified only by the JHM are also included in the spreadsheet document provided in the on-line supporting materials).  

\begin{figure}[h!]
  \centering
\includegraphics[width=15cm]{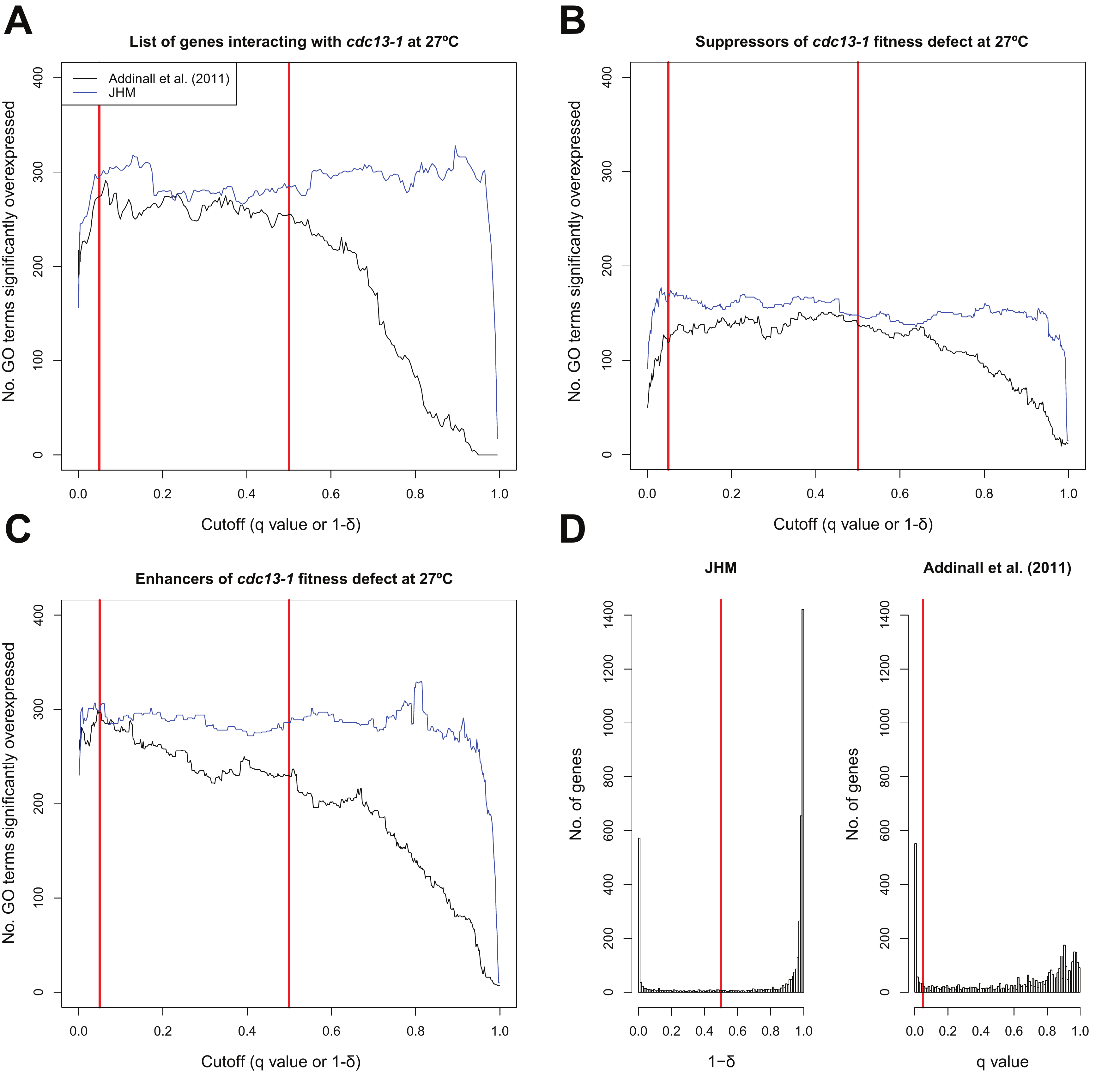}
\caption{
Sensitivity to significance thresholds.
A), B) \& C)  Comparison of the number of significantly over-expressed GO terms identified in lists of significant interactors found using the \citet{QFA1} method and using the JHM.
D)  Frequency histograms showing distributions of classifier values after looking for genes interacting with \emph{cdc13-1} at 27C using the Addinall et al. (2011) method or using the JHM.
}
\label{fig:specsens}
\end{figure}

A major advantage of the Bayesian approaches presented here over the \citet{QFA1} approach is the measure used for classifying significant interactions.  Classifying interactions with a posterior estimate for $\delta_{l}$ (the probability that an interaction exists) greater than $0.5$ as significant is less arbitrary than the traditional frequentist approach of classifying interactions with p-values less than $0.05$ as significant.
Examining how the number of over-represented GO terms found in lists of interactors varies with the classification threshold shows that the Bayesian JHM approach is also less sensitive to the precise threshold values used.
Figure \ref{fig:specsens} shows that the number of over-expressed GO terms found among hits is relatively stable in the region of $\delta_{l}=0.5$ for the JHM compared to the equivalent number in the region of $q=0.05$ for the \citet{QFA1} approach.
Significantly over-expressed GO terms were identified using the \texttt{hyperGTest} function in the GOstats R package.
Note that the values used to classify whether a gene interacts with \emph{cdc13-1} at 27C (q-value and $\delta$ respectively, red vertical lines as presented in Section~4.4) are not directly comparable; however the full range of possible cutoffs for both values are plotted.
In particular, using the frequentist \citet{QFA1} approach, the number of over-expressed GO terms falls rapidly where $q<0.05$.
We tested whether this observation depended on our choice of the parameter $p$, which represents our prior expectation of the proportion of genes interacting with the query, by generating similar sensitivity plots for $p$ between 0.01 and 0.2 (Section~6 of the on-line supporting materials).
We observed similar profiles of over-expressed GO terms for all values of $p$ tested.

Comparing the genetic interaction strengths generated by the Bayesian hierarchical models and frequentist analysis, we find that the results are largely similar (Section~6 of the on-line supporting materials); however the GO term analysis described above suggests that the differences are important.

The results of a simulation study comparing the sensitivity and specificity of the \citet{QFA1} approach, the REM, the SHM and the JHM are summarised in section~8 of the supplementary materials.  We find that the JHM correctly identified a higher proportion of ``true'' interactions in a synthetic dataset than the \citet{QFA1} approach.

\subsubsection{\label{hierarchy}Hierarchy and model parameters}
The hierarchical structure and model choices included in the Bayesian JHM and IHM are derived from the known experimental structure of QFA.
Different levels of variation for different $\emph{orf}\Delta$s are expected and can be observed by comparing distributions of frequentist estimates or by visual inspection of yeast culture images.
The direct relationship between experimental and model structure, together with the richness of detail and number of replicates included in QFA experimental design, reassures us that overfitting is not an issue in this analysis.
For the \emph{ura3$\Delta$}~${{27}^{\circ}}$C and \mbox{\emph{cdc13-1}}~${27}^{{\circ}}$C experiment with $\sim$4,294 $\emph{orf}\Delta$s there are $\sim$1.25 times the number of parameters in the JHM ($\sim$200,000) compared to the two stage REM approach ($\sim$160,000) but when compared to the large number of pairs of data points ($\sim$830,000) there are sufficient degrees of freedom to justify our proposed Bayesian models.

\subsubsection{\label{computing_reqs}Computing requirements}
Our Bayesian hierarchical models require significant computational time. 
As expected, the mixing of chains in our models is weakest at population level parameters such as $K_p$ and $\alpha_c$.
For the \emph{ura3$\Delta$} ${{27}^{\circ}}$C and \mbox{\emph{cdc13-1}} ${27}^{{\circ}}$C \hl{dataset}, running with an MCMC burn-in of 800,000 updates, followed by generating 1,000 samples thinned by a factor of 100, the JHM takes ${\sim}4$ weeks to converge and produce a sufficiently large sample.  The two stage Bayesian approach takes one week (with the IHM part taking ${\sim}1$ day), whereas the REM takes ${\sim}3$ days and the \cite{QFA1} approach takes ${\sim}3$ hours. 
A QFA experiment can take over a month from start to finish and so analysis time is acceptable in comparison to the time taken for the creation of the data set but still a notable inconvenience.
We expect that with further research effort, computational time can be decreased by using an improved inference scheme and that inference for the JHM could be completed in less than a week without parallelisation.
 
  \section{\label{sec:Discussion}\rev{Discussion}}

We have joined a hierarchical model of microbial growth with a model for genetic interaction in order to learn about strain fitnesses, evidence for genetic interaction and interaction strengths simultaneously. 
By introducing Bayesian methodology to QFA we have been able to model the hierarchical nature of the experiment and expand the multiplicative model for genetic interaction to incorporate many sources of variation that previously had to be ignored.

We propose two new Bayesian hierarchical models to replace the current statistical analysis for identifying genetic interactions within a QFA screen comparison. 
The two-stage approach fits the SHM followed by the IHM, \hl{with univariate point estimate fitness} definitions \hl{generated} as an intermediate step. 
The two-stage approach can therefore be regarded as a Bayesian hierarchical version of the \cite{QFA1} approach.
\hl{In contrast,} the one-stage approach fits the JHM, which does not require a \hl{separate definition of} fitness, allowing interaction to be identified \hl{by either growth rate (logistic parameter $r$) or final biomass achievable (logistic parameter $K$) by a given genotype}. 
Our one-stage approach is a new method for detecting genetic interaction that further develops the interpretation of epistasis within QFA screens.
 
We present a hierarchical, frequentist approach using random effects, namely the REM, in an attempt to improve on the \cite{QFA1} approach. Due to the lack of flexibility in modelling assumptions allowable, the REM is \hl{unsuitable for modelling} the distribution of \rev{\emph{orf}$\Delta$ level variation} or for simultaneously modelling genetic interaction and logistic growth curves.  

\rev{The data from which logistic parameter estimates are derived during QFA are the result of a technically challenging, high-throughput experimental procedure with a diverse range of possible technical errors.  Our Bayesian, hierarchical models allow us the flexibility to make distributional assumptions that more closely match the data.
This allows us to switch between modelling parameter uncertainty with Normal, Log-Normal and Student's $t$ distribution where appropriate.
}  

QFA experimental design is intrinsically multilevel and is therefore more closely modelled by our hierarchical scheme. Consequently the JHM and IHM capture sources of variation not considered by \cite{QFA1}. By sharing information across levels in the hierarchy, our models have allowed us to learn more about $\emph{orf}\Delta$s with weaker genetic interaction.
Our more flexible model of variance also avoids misclassification of individual genotypes with high variance as having significant interactions.
Without fully accounting for the variation described in the Bayesian hierarchical models, the previous \cite{QFA1} approach may have relatively poor power to detect subtle interactions, obscuring potential novel observations.

Many subtle, interesting genetic interactions may remain to be identified in the data from the QFA experiments we re-analyse in this paper.
The JHM is better able to identify subtle interactions.
For example, strains with little evidence for interaction with a background mutation in terms of growth rate but with strong evidence of interaction in terms of carrying capacity are sometimes classified as interactors using the JHM (see Figure~\ref{fig:JHM_only}).
In our two-stage approaches, univariate fitness measures such as $D_R\times D_P$ are used in the intermediate steps, occasionally causing interaction in terms of one parameter to be masked by the other. 

\hl{As expected, many genes previously unidentified by \cite{QFA1} have been identified as showing evidence of interaction using both of our Bayesian hierarchical modelling approaches.}
\rev{Genes which have been identified only by the JHM (see Figure~\ref{fig:old}D), such as those showing interaction only in terms of $r$, are found to be related to telomere biology in the literature.}
\rev{Currently there is not sufficient information available to identify the proportion of identified interactions that are true hits and so we use unbiased GO term enrichment analyses to confirm that the lists of genetic interactions closely reflect the true underlying biology.
GO term annotations relevant to telomere biology are available for well-studied genes in the current literature. Unsurprisingly all of the approaches considered closely reflect the most well-known GO terms (see Table~\ref{tab:sup_enh}).}

Computational time for the new Bayesian approach ranges from one to four weeks for one of the datasets presented in \cite{QFA1}. This is of the same magnitude as the time taken to design and execute the experimental component of QFA (approximately six weeks). 

Overall \hl{we recommend} a JHM or ``Bayesian QFA'' for analysis of current and future QFA data sets as it accounts for more sources of variation than the \cite{QFA1} QFA methodology.
With the JHM we have outlined new genes with significant evidence of interaction in the \emph{ura3}$\Delta$~${{27}^{\circ}}$C~and~\mbox{\emph{cdc13-1}}~${27}^{{\circ}}$C experiment.
The new Bayesian hierarchical models we present here will also be suitable for identifying new genes showing evidence of genetic interaction in backgrounds other than telomere activity.
\hl{We hope that further, reductionist} lab work by experimental biologists will give additional insight into the mechanisms by which the new genes we have uncovered interact with the telomere. 
 \section*{Acknowledgements}
This research was supported by grants from the Biotechnology and Biological Sciences Research Council, UK (BBSRC): BBF016980/1, the Medical Research Council, UK (MRC): MR/L001284/1 and the Wellcome Trust: 075294, 093088.  The authors would like to thank the editor and anonymous reviewers for their valuable comments and suggestions.\vspace*{-8pt}
 \bibliography{references}             

\begin{thebibliography}{}

\bibitem[\protect\citeauthoryear{Addinall, Downey, Yu, Zubko, Dewar, Leake,
  Hallinan, Shaw, James, Wilkinson, Wipat, Durocher, and Lydall}{Addinall
  et~al.}{2008}]{RTC6}
Addinall, S.~G., M.~Downey, M.~Yu, M.~K. Zubko, J.~Dewar, A.~Leake,
  J.~Hallinan, O.~Shaw, K.~James, D.~J. Wilkinson, A.~Wipat, D.~Durocher, and
  D.~Lydall (2008, Dec).
\newblock {{A} genomewide suppressor and enhancer analysis of cdc13-1 reveals
  varied cellular processes influencing telomere capping in {S}accharomyces
  cerevisiae}.
\newblock {\em Genetics\/}~{\em 180\/}(4), 2251--2266.

\bibitem[\protect\citeauthoryear{Addinall, Holstein, Lawless, Yu, Chapman,
  Banks, Ngo, Maringele, Taschuk, Young, Ciesiolka, Lister, Wipat, Wilkinson,
  and Lydall}{Addinall et~al.}{2011}]{QFA1}
Addinall, S.~G., E.-M. Holstein, C.~Lawless, M.~Yu, K.~Chapman, A.~P. Banks,
  H.-P. Ngo, L.~Maringele, M.~Taschuk, A.~Young, A.~Ciesiolka, A.~L. Lister,
  A.~Wipat, D.~J. Wilkinson, and D.~Lydall (2011).
\newblock {Quantitative Fitness Analysis Shows That NMD Proteins and Many Other
  Protein Complexes Suppress or Enhance Distinct Telomere Cap Defects}.
\newblock {\em PLoS Genet\/}~{\em 7\/}(4), e1001362.

\bibitem[\protect\citeauthoryear{Aylor and Zeng}{Aylor and Zeng}{2008}]{epis3}
Aylor, D.~L. and Z.-B. Zeng (2008).
\newblock From classical genetics to quantitative genetics to systems biology:
  Modeling epistasis.
\newblock {\em PLoS Genet\/}~{\em 4\/}(3), e1000029.

\bibitem[\protect\citeauthoryear{Banks, Lawless, and Lydall}{Banks
  et~al.}{2012}]{jove}
Banks, A., C.~Lawless, and D.~Lydall (2012).
\newblock {A Quantitative Fitness Analysis Workflow}.
\newblock {\em J. Vis. Exp\/}~{\em 66}, e4018.

\bibitem[\protect\citeauthoryear{Benjamini and Hochberg}{Benjamini and
  Hochberg}{1995}]{ben_hoc}
Benjamini, Y. and Y.~Hochberg (1995).
\newblock {Controlling the False Discovery Rate: A Practical and Powerful
  Approach to Multiple Testing}.
\newblock {\em Journal of the Royal Statistical Society. Series B
  (Methodological)\/}~{\em 57\/}(1), 289--300.

\bibitem[\protect\citeauthoryear{Bernardo and Smith}{Bernardo and
  Smith}{2007}]{Bayth}
Bernardo, J. and A.~Smith (2007).
\newblock {\em Bayesian Theory}.
\newblock Wiley Series in Probability and Statistics. John Wiley \& Sons
  Canada, Limited.

\bibitem[\protect\citeauthoryear{Cordell}{Cordell}{2002}]{cordell2002epistasis}
Cordell, H.~J. (2002).
\newblock Epistasis: what it means, what it doesn't mean, and statistical
  methods to detect it in humans.
\newblock {\em Human molecular genetics\/}~{\em 11\/}(20), 2463--2468.

\bibitem[\protect\citeauthoryear{Falcon and Gentleman}{Falcon and
  Gentleman}{2007}]{GOstats}
Falcon, S. and R.~Gentleman (2007).
\newblock Using {G}{O}stats to test gene lists for {G}{O} term association.
\newblock {\em Bioinformatics\/}~{\em 23\/}(2), 257--8.

\bibitem[\protect\citeauthoryear{Foster, Zubko, Guillard, and Lydall}{Foster
  et~al.}{2006}]{MRX}
Foster, S.~S., M.~K. Zubko, S.~Guillard, and D.~Lydall (2006).
\newblock {MRX protects telomeric DNA at uncapped telomeres of budding yeast
  \emph{cdc13-1} mutants}.
\newblock {\em DNA Repair\/}~{\em 5\/}(7), 840 -- 851.

\bibitem[\protect\citeauthoryear{Franke, Gehlen, and Ehrenhofer-Murray}{Franke
  et~al.}{2008}]{TGS1}
Franke, J., J.~Gehlen, and A.~E. Ehrenhofer-Murray (2008, Nov).
\newblock {{H}ypermethylation of yeast telomerase {R}{N}{A} by the sn{R}{N}{A}
  and sno{R}{N}{A} methyltransferase {T}gs1}.
\newblock {\em J. Cell. Sci.\/}~{\em 121\/}(Pt 21), 3553--3560.

\bibitem[\protect\citeauthoryear{Gelman}{Gelman}{2006}]{Gelmanprior}
Gelman, A. (2006).
\newblock Prior distributions for variance parameters in hierarchical models.
\newblock {\em Bayesian analysis\/}~{\em 1\/}(3), 515--533.

\bibitem[\protect\citeauthoryear{Gelman and Hill}{Gelman and
  Hill}{2006}]{GelmanMultilevel}
Gelman, A. and J.~Hill (2006).
\newblock {\em {Data Analysis Using Regression and Multilevel/Hierarchical
  Models}\/} (1 ed.).
\newblock Cambridge University Press.

\bibitem[\protect\citeauthoryear{Goldstein}{Goldstein}{2011}]{BayHi}
Goldstein, H. (2011).
\newblock {\em Multilevel Statistical Models}.
\newblock Wiley Series in Probability and Statistics. Wiley.

\bibitem[\protect\citeauthoryear{Greider and Blackburn}{Greider and
  Blackburn}{1985}]{greider1985identification}
Greider, C.~W. and E.~H. Blackburn (1985).
\newblock Identification of a specific telomere terminal transferase activity
  in tetrahymena extracts.
\newblock {\em cell\/}~{\em 43\/}(2), 405--413.

\bibitem[\protect\citeauthoryear{Heidelberger and Welch}{Heidelberger and
  Welch}{1981}]{Heidelberger}
Heidelberger, P. and P.~D. Welch (1981).
\newblock {A spectral method for confidence interval generation and run length
  control in simulations}.
\newblock {\em Commun. ACM\/}~{\em 24\/}(4), 233--245.

\bibitem[\protect\citeauthoryear{{Jow}, {Boys}, and {Wilkinson}}{{Jow}
  et~al.}{2014}]{Jow2014}
{Jow}, H., R.~J. {Boys}, and D.~J. {Wilkinson} (2014, July).
\newblock {Bayesian identification of protein differential expression in
  multi-group isobaric labelled mass spectrometry data}.
\newblock {\em ArXiv e-prints\/}.

\bibitem[\protect\citeauthoryear{Keogh, Kurdistani, Morris, Ahn, Podolny,
  Collins, Schuldiner, Chin, Punna, Thompson, Boone, Emili, Weissman, Hughes,
  Strahl, Grunstein, Greenblatt, Buratowski, and Krogan}{Keogh
  et~al.}{2005}]{CTI6}
Keogh, M.~C., S.~K. Kurdistani, S.~A. Morris, S.~H. Ahn, V.~Podolny, S.~R.
  Collins, M.~Schuldiner, K.~Chin, T.~Punna, N.~J. Thompson, C.~Boone,
  A.~Emili, J.~S. Weissman, T.~R. Hughes, B.~D. Strahl, M.~Grunstein, J.~F.
  Greenblatt, S.~Buratowski, and N.~J. Krogan (2005, Nov).
\newblock {{C}otranscriptional set2 methylation of histone {H}3 lysine 36
  recruits a repressive {R}pd3 complex}.
\newblock {\em Cell\/}~{\em 123\/}(4), 593--605.

\bibitem[\protect\citeauthoryear{Lawless, Wilkinson, Young, Addinall, and
  Lydall}{Lawless et~al.}{2010}]{Colonyzer}
Lawless, C., D.~J. Wilkinson, A.~Young, S.~G. Addinall, and D.~A. Lydall
  (2010).
\newblock {Colonyzer: automated quantification of micro-organism growth
  characteristics on solid agar}.
\newblock {\em BMC Bioinformatics\/}~{\em 11}, 287.

\bibitem[\protect\citeauthoryear{Mani, St.Onge, Hartman, Giaever, and
  Roth}{Mani et~al.}{2008}]{epis2}
Mani, R., R.~P. St.Onge, J.~L. Hartman, G.~Giaever, and F.~P. Roth (2008).
\newblock {Defining genetic interaction}.
\newblock {\em Proceedings of the National Academy of Sciences\/}~{\em
  105\/}(9), 3461--3466.

\bibitem[\protect\citeauthoryear{Nugent, Hughes, Lue, and Lundblad}{Nugent
  et~al.}{1996}]{cdc131}
Nugent, C.~I., T.~R. Hughes, N.~F. Lue, and V.~Lundblad (1996).
\newblock {Cdc13p: A Single-Strand Telomeric DNA-Binding Protein with a Dual
  Role in Yeast Telomere Maintenance}.
\newblock {\em Science\/}~{\em 274\/}(5285), 249--252.

\bibitem[\protect\citeauthoryear{O'Hara and Sillanpaa}{O'Hara and
  Sillanpaa}{2009}]{indicator}
O'Hara, R.~B. and M.~J. Sillanpaa (2009).
\newblock {A Review of Bayesian Variable Selection Methods: What, How and
  Which}.
\newblock {\em Bayesian Analysis\/}~{\em 4}, 85.

\bibitem[\protect\citeauthoryear{Olovnikov}{Olovnikov}{1973}]{telo}
Olovnikov, A. (1973, sep).
\newblock A theory of marginotomy.
\newblock {\em Journal of Theoretical Biology\/}~{\em 41\/}(1), 181--190.

\bibitem[\protect\citeauthoryear{Phenix, Morin, Batenchuk, Parker, Abedi, Yang,
  Tepliakova, Perkins, and Kærn}{Phenix et~al.}{2011}]{epis1}
Phenix, H., K.~Morin, C.~Batenchuk, J.~Parker, V.~Abedi, L.~Yang,
  L.~Tepliakova, T.~J. Perkins, and M.~Kærn (2011).
\newblock {Quantitative Epistasis Analysis and Pathway Inference from Genetic
  Interaction Data}.
\newblock {\em PLoS Comput Biol\/}~{\em 7\/}(5), e1002048.

\bibitem[\protect\citeauthoryear{Phillips}{Phillips}{1998}]{epis4}
Phillips, P.~C. (1998).
\newblock The language of gene interaction.
\newblock {\em Genetics\/}~{\em 149\/}(3), 1167--1171.

\bibitem[\protect\citeauthoryear{Pinheiro and Bates}{Pinheiro and
  Bates}{2000}]{nlme}
Pinheiro, J.~C. and D.~M. Bates (2000).
\newblock {\em {Mixed Effects Models in S and S-Plus}}.
\newblock Springer.

\bibitem[\protect\citeauthoryear{Raftery and Lewis}{Raftery and
  Lewis}{1995}]{Raftery}
Raftery, A.~E. and S.~M. Lewis (1995).
\newblock {The Number of Iterations, Convergence Diagnostics and Generic
  Metropolis Algorithms}.
\newblock In {\em {In Practical Markov Chain Monte Carlo (W.R. Gilks, D.J.
  Spiegelhalter and S. Richardson, eds.)}}, pp.\  115--130. Chapman and Hall.

\bibitem[\protect\citeauthoryear{Schuldiner, Collins, Weissman, and
  Krogan}{Schuldiner et~al.}{2006}]{emap}
Schuldiner, M., S.~Collins, J.~Weissman, and N.~Krogan (2006).
\newblock {Quantitative genetic analysis in Saccharomyces cerevisiae using
  epistatic miniarray profiles (E-MAPs) and its application to chromatin
  functions}.
\newblock {\em Methods\/}~{\em 40\/}(4), 344 -- 352.

\bibitem[\protect\citeauthoryear{Tong and Boone}{Tong and
  Boone}{2006}]{sgaboone}
Tong, A.~H. and C.~Boone (2006).
\newblock {Synthetic genetic array analysis in Saccharomyces cerevisiae.}
\newblock {\em Methods Mol Biol\/}~{\em 313}, 171--192.

\bibitem[\protect\citeauthoryear{Verhulst}{Verhulst}{1845}]{Verhulst1847}
Verhulst, P.~F. (1845).
\newblock {Recherches math\'{e}matiques sur la loi d'accroissement de la
  population.}
\newblock {\em Nouveaux m\'{e}moires de l'Academie Royale des Science et
  Belles-Lettres de Bruxelles\/}~{\em 18}, 1--41.

\bibitem[\protect\citeauthoryear{Yi}{Yi}{2010}]{hierarchical1}
Yi, N. (2010).
\newblock {Statistical analysis of genetic interactions.}
\newblock {\em Genetics research\/}~{\em 92\/}(5-6), 443--459.

\bibitem[\protect\citeauthoryear{Zhang, Baladandayuthapani, Mallick, Manyam,
  Thompson, Bondy, and Do}{Zhang et~al.}{2014}]{Zhang2014}
Zhang, L., V.~Baladandayuthapani, B.~K. Mallick, G.~C. Manyam, P.~A. Thompson,
  M.~L. Bondy, and K.-A. Do (2014).
\newblock Bayesian hierarchical structured variable selection methods with
  application to molecular inversion probe studies in breast cancer.
\newblock {\em Journal of the Royal Statistical Society: Series C (Applied
  Statistics)\/}~{\em 63\/}(4), 595--620.

\end{thebibliography}
  \bibliographystyle{Chicago}

\end{document}


\begin{center}{\bf Web-based supporting materials for ``Bayesian hierarchical modelling for inferring genetic interactions in yeast'' by Jonathan Heydari, Conor Lawless, David A. Lydall and Darren J. Wilkinson}
\end{center}

\setcounter{figure}{0}
\renewcommand\thefigure{\thesection.\arabic{figure}}    
\setcounter{table}{0}
\renewcommand\thetable{\thesection.\arabic{table}}    

\section{\label{app:plate} Plate diagrams}
\begin{figure}[H]
\centering
\makebox{\includegraphics[scale=0.63]{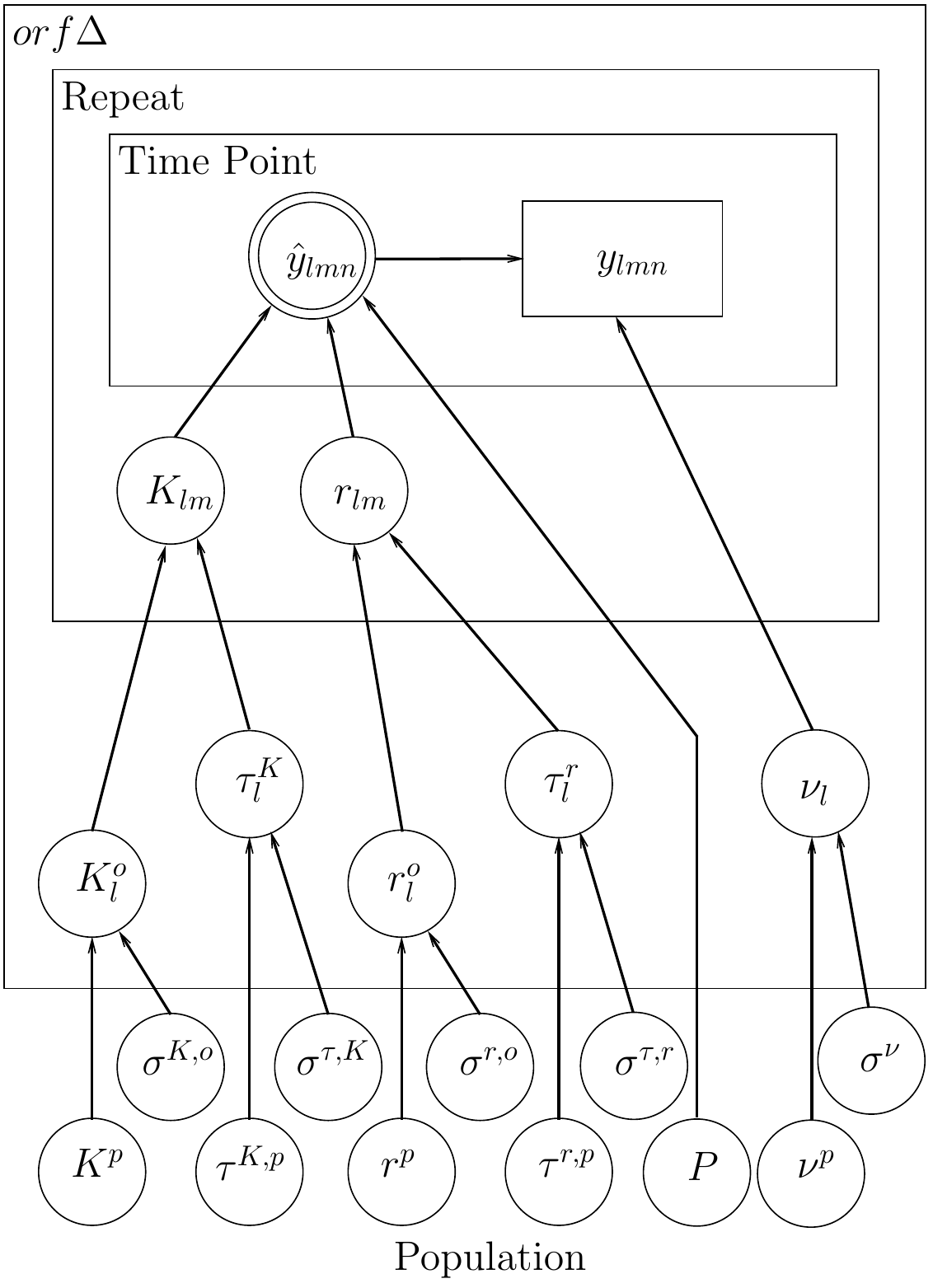}}
\caption[Directed acyclic graph for the separate hierarchical model]{\hl{Plate diagram} for the SHM, described in Section~3.1 of the main article. This figure shows the four levels of hierarchy in the SHM model, population, $\emph{orf}\Delta$ ($l$), repeat ($m$) and time point ($n$).
Prior hyperparameters for the population parameters are omitted.
A circular node represents a parameter in the model. 
An arrow from a source node to a target node indicates that the source node parameter is a \hl{prior hyperparameter} for the target node parameter. 
Each rectangular box corresponds to a level of the hierarchy. 
Nodes within multiple boxes are nested and their parameters are indexed by corresponding levels of the hierarchy. The node consisting of two concentric circles corresponds to our model's fitted values. 
The rectangular node represents the observed \hl{data}.}
\label{fig:SHMDAG}
\end{figure}

\begin{figure}[h!]
\centering
\makebox{\includegraphics[scale=0.65]{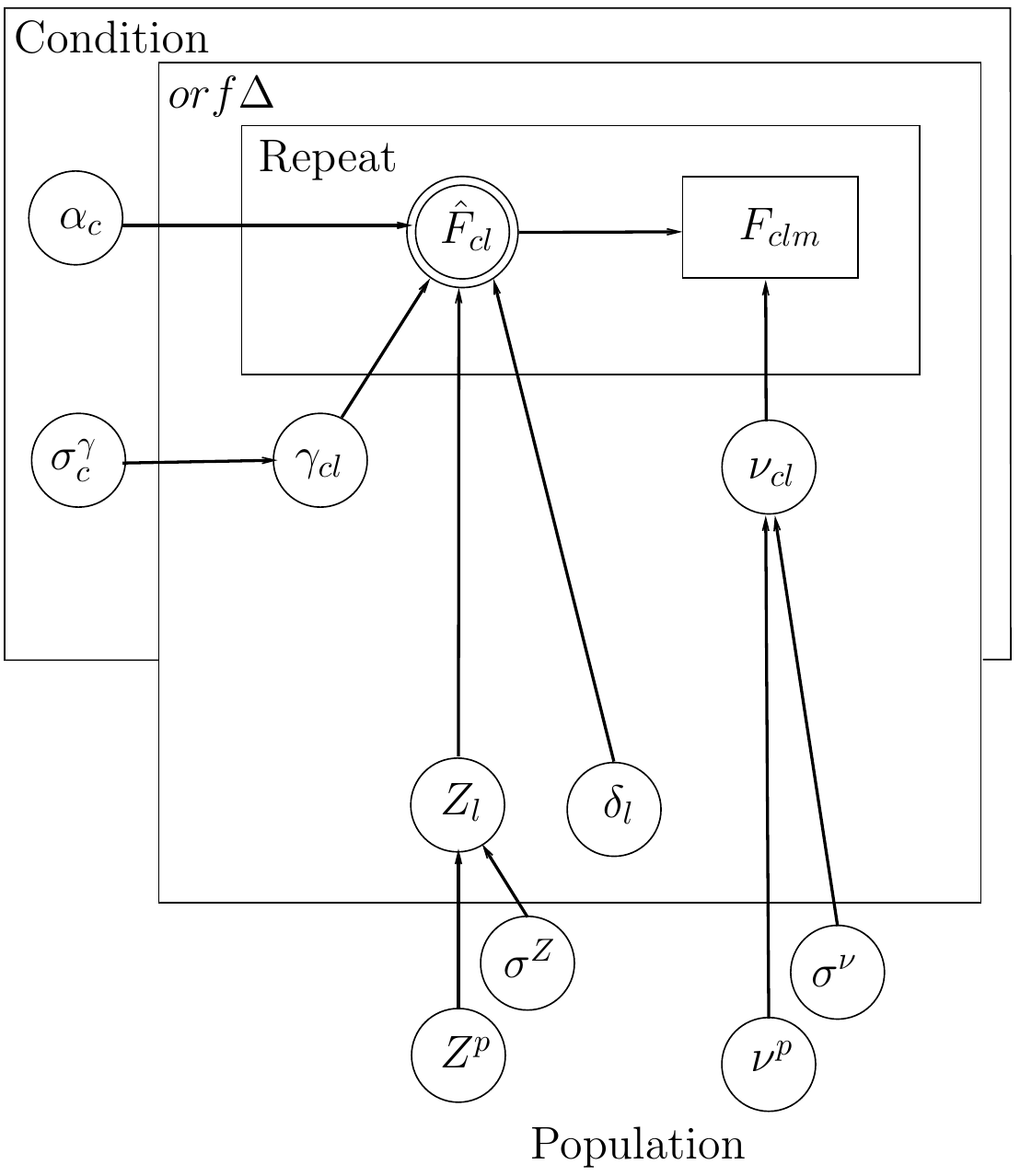}}
\caption{\hl{Plate diagram} for the IHM, described in Section~3.2 of the main article. 
This figure shows the four levels of hierarchy in the IHM model: population, $\emph{orf}\Delta$ ($l$), condition ($c$) and repeat ($m$).
Prior hyperparameters for population parameters are omitted.
Plate diagram notation as in Figure~\ref{fig:SHMDAG}.
}
\label{fig:IHMDAG}
\end{figure}

\begin{figure}[h!]
\centering
\makebox{\includegraphics[scale=0.65]{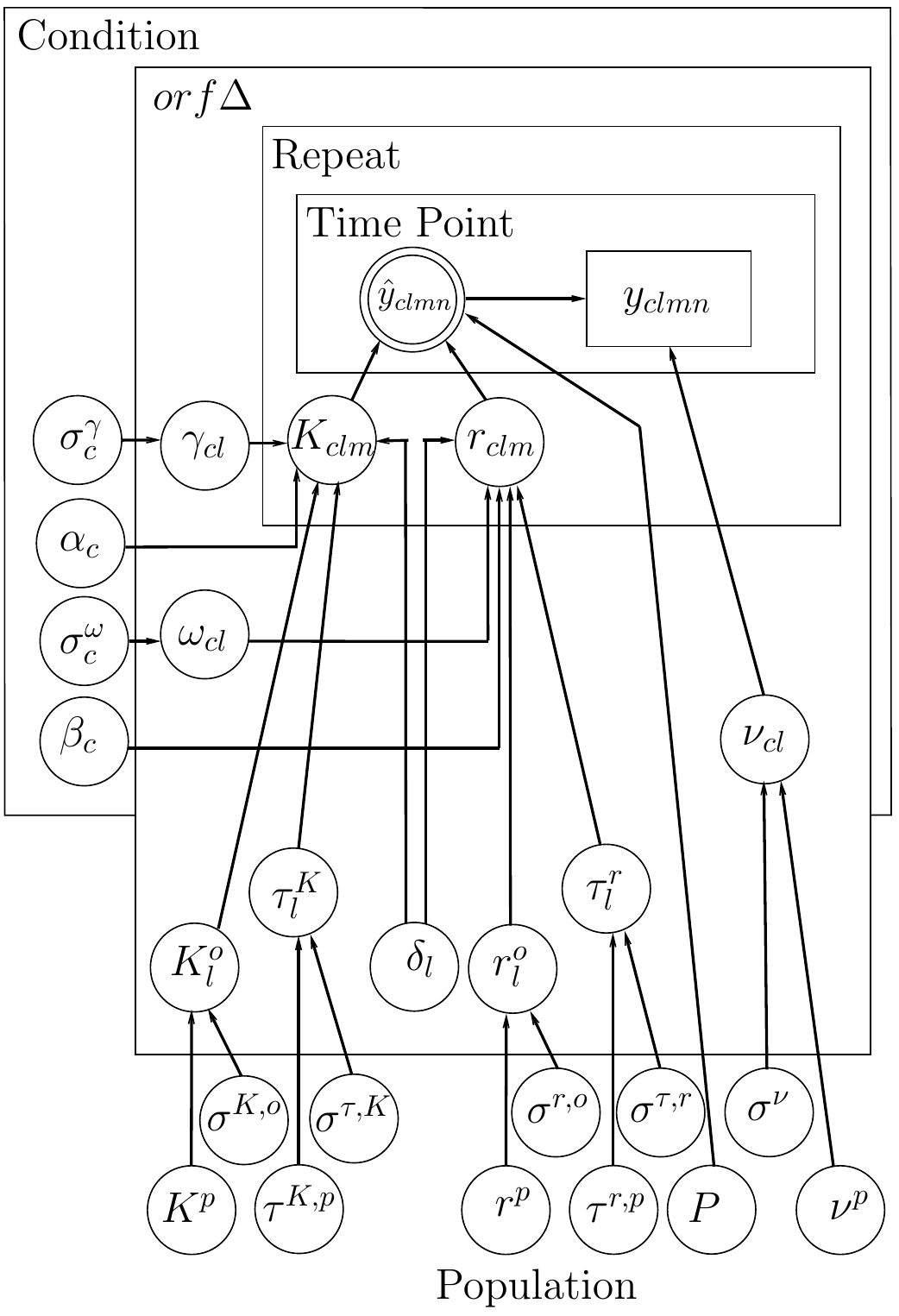}}
\caption{\hl{Plate diagram} the JHM, described in Section~3.3 of the main article. 
This figure shows the five levels of hierarchy in the JHM model, population, $\emph{orf}\Delta$ ($l$), condition ($c$), repeat ($m$) and time point ($n$). 
Prior hyperparameters for the population parameters are omitted.
Plate diagram notation is given in Figure~\ref{fig:SHMDAG}.
}
\label{fig:JHMDAG}
\end{figure}
\clearpage

\setcounter{figure}{0}
\setcounter{table}{0}
\section{GO term enrichment analysis in R}
{\fontsize{9}{9}\selectfont
\begin{verbatim}
source("http://bioconductor.org/biocLite.R")
biocLite("GOstats")
biocLite("org.Sc.sgd.db")
###################
library(GOstats) # GO testing tool package
library(org.Sc.sgd.db) # yeast gene annotation package
genes=read.table("sm_JHM_list.txt", header=T)
UNIVSTRIP=genes[,2]
genes<-as.vector(genes[genes[,3]>0.5,2])
genes<-unique(genes)
ensemblIDs=as.list(org.Sc.sgdPMID2ORF)
univ=unlist(ensemblIDs)
univ=univ[!is.na(univ)]
length(univ)
length(unique(univ))
univ=unique(univ)
all=as.vector(univ)
all=all[all%in%UNIVSTRIP]
length(all)
ontology=c("BP")
vec<-genes%in%univ
genes<-genes[vec]
params_temp=new("GOHyperGParams", geneIds=genes, 
universeGeneIds=all,
 annotation="org.Sc.sgd.db", categoryName="GO",
 ontology=ontology, pvalueCutoff=1, 
 testDirection = "over")
results=hyperGTest(params_temp)
results=summary(results)
results$qvalue<-p.adjust(results$Pvalue,method="BH")
\end{verbatim}
}

\clearpage
\section{\label{app:interactions}Fitness plots with GO terms highlighted}
\setcounter{figure}{0}
\setcounter{table}{0}
\begin{figure}[h!]
  \centering
\includegraphics[width=14cm]{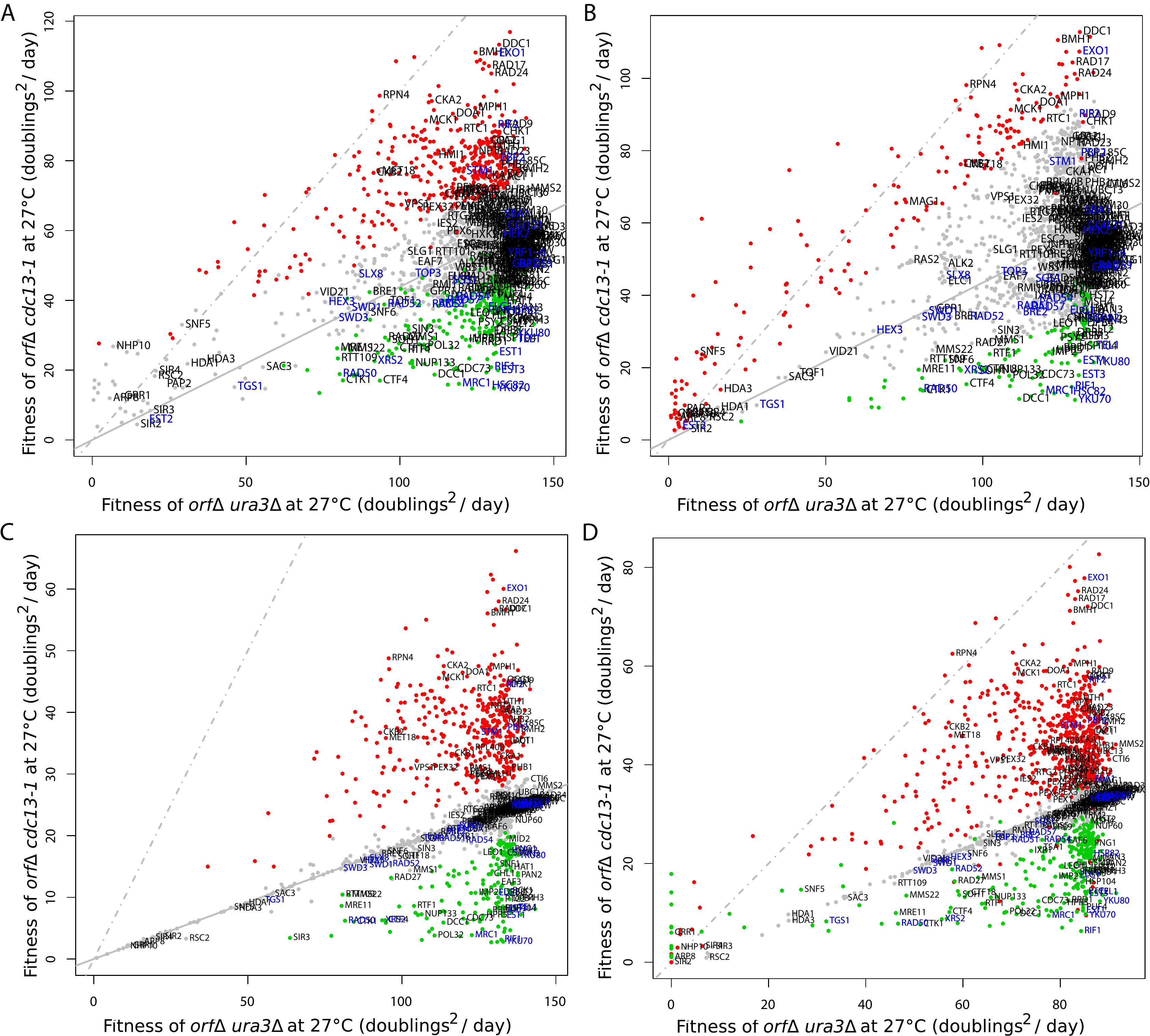}
\caption{Alternative fitness plots with \emph{orf}$\Delta$ posterior mean fitnesses. Text for the ``telomere maintenance'' GO term is highlighted in blue.
A) Non-Bayesian, non-hierarchical fitness plot, based on Table~S6 from Addinall et al. (2011) $(F=MDR\times MDP)$.
B) Non-Bayesian, hierarchical fitness plot, \hl{from fitting REM to data} in Table~S6 from Addinall et al. (2011) $(F=MDR\times MDP)$.
C) IHM fitness plot with $\emph{orf}\Delta$ posterior mean fitness $(F=MDR\times MDP)$.
D) JHM fitness plot with $\emph{orf}\Delta$ posterior mean fitnesses.
$\emph{orf}\Delta$ strains are classified as being a suppressor or enhancer based on analysis of growth parameter $r$.
Further explanation and notation for fitness plots are given in Figure~3 of the main article.
}
\end{figure}
\begin{figure}[h!]
  \centering
\includegraphics[width=14cm]{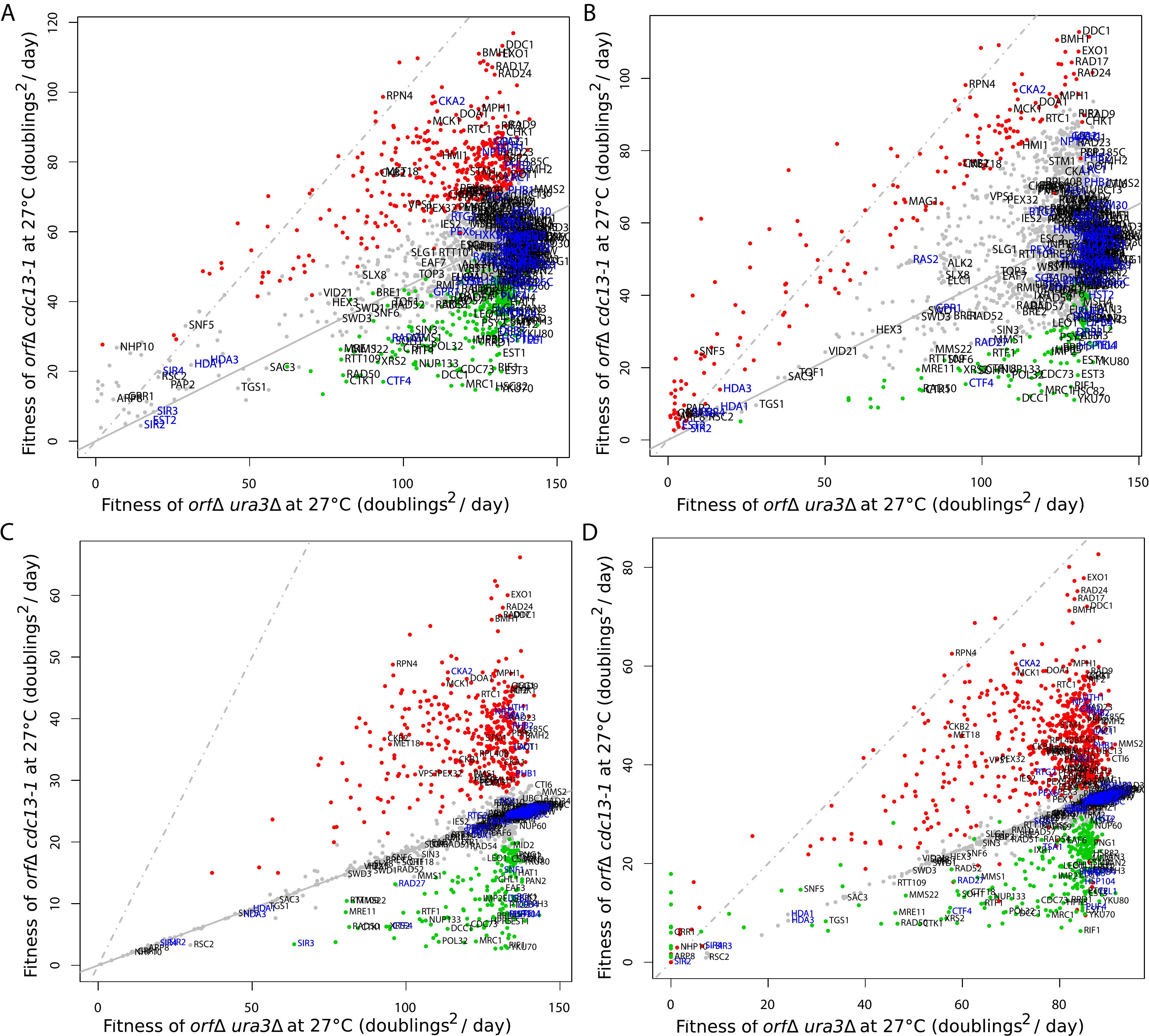}
\caption{Alternative fitness plots with $\emph{orf}\Delta$ posterior mean fitnesses. Text for the ``ageing'' GO term is highlighted in blue.
A) Non-Bayesian, non-hierarchical fitness plot, based on Table~S6 from Addinall et al. (2011) $(F=MDR\times MDP)$.
B) Non-Bayesian, hierarchical fitness plot, \hl{from fitting REM to data} in Table~S6 from Addinall et al. (2011) $(F=MDR\times MDP)$.
C) IHM fitness plot with $\emph{orf}\Delta$ posterior mean fitness $(F=MDR\times MDP)$.
D) JHM fitness plot with $\emph{orf}\Delta$ posterior mean fitnesses.
$\emph{orf}\Delta$ strains are classified as being a suppressor or enhancer based on analysis of growth parameter $r$.
Further explanation and notation for fitness plots are given in Figure~3 of the main article.
}
\end{figure}
\begin{figure}[h!]
  \centering
\includegraphics[width=14cm]{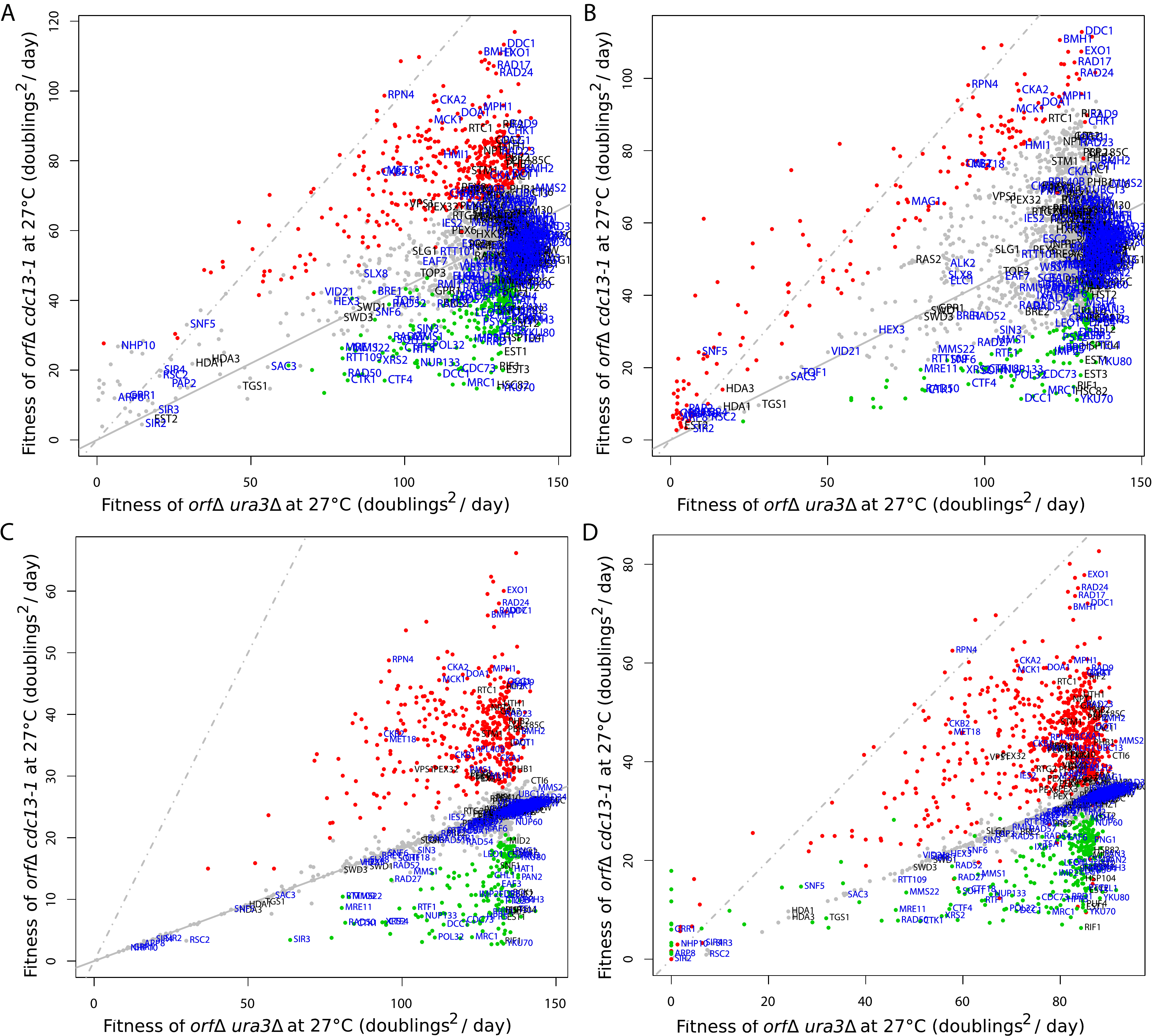}
\caption{Alternative fitness plots with $\emph{orf}\Delta$ posterior mean fitnesses. Text for the ``response to DNA damage'' GO term is highlighted in blue.
A) Non-Bayesian, non-hierarchical fitness plot, based on Table~S6 from Addinall et al. (2011) $(F=MDR\times MDP)$.
B) Non-Bayesian, hierarchical fitness plot, \hl{from fitting REM to data} in Table~S6 from Addinall et al. (2011) $(F=MDR\times MDP)$.
C) IHM fitness plot with $\emph{orf}\Delta$ posterior mean fitness $(F=MDR\times MDP)$.
D) JHM fitness plot with $\emph{orf}\Delta$ posterior mean fitnesses.
$\emph{orf}\Delta$ strains are classified as being a suppressor or enhancer based on analysis of growth parameter $r$.
Further explanation and notation for fitness plots are given in Figure~3 of the main article.
}
\end{figure}
\begin{figure}[h!]
  \centering
\includegraphics[width=14cm]{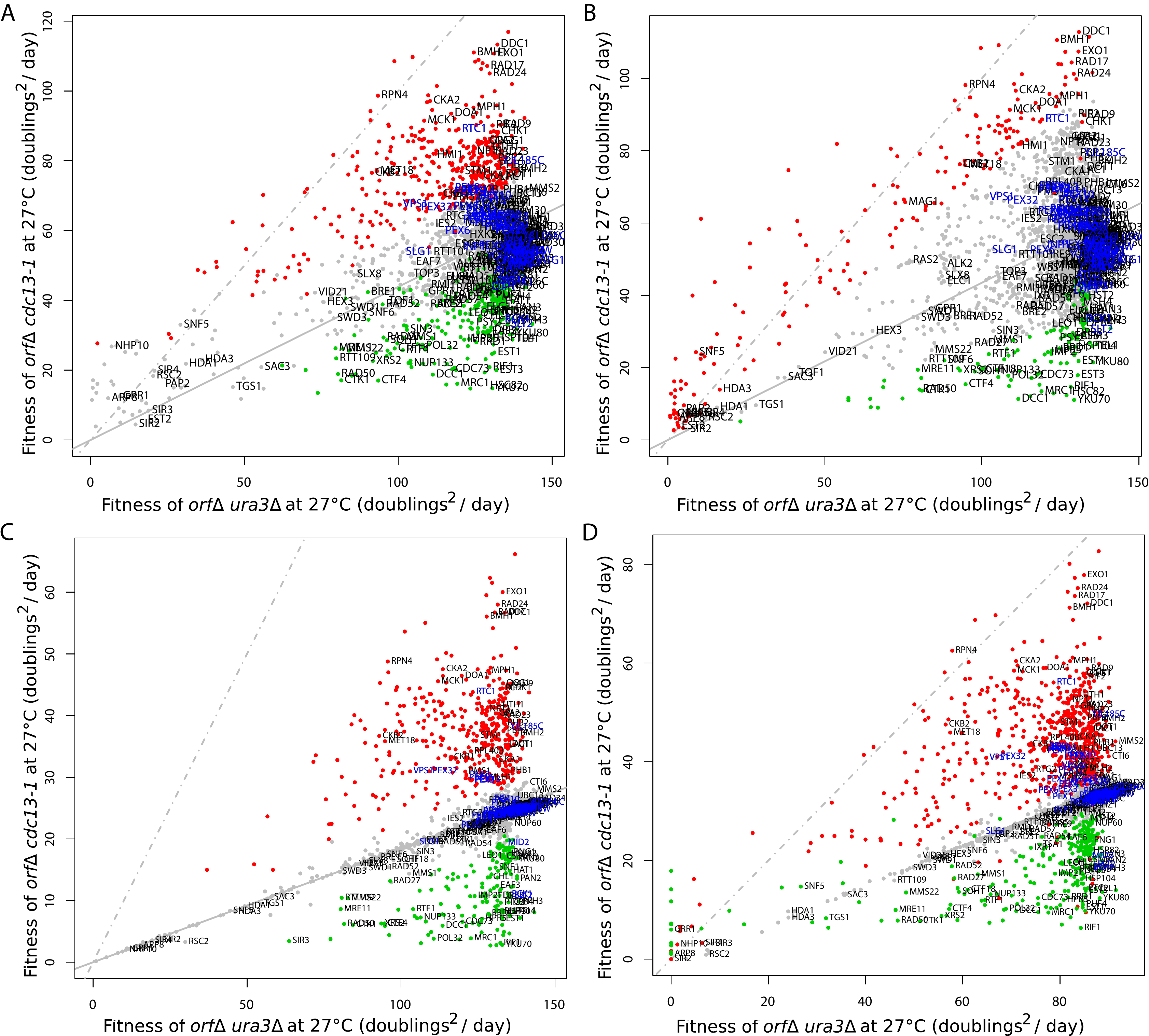}
\caption{Alternative fitness plots with $\emph{orf}\Delta$ posterior mean fitnesses. Text for the ``peroxisomal organisation'' GO term is highlighted in blue.
A) Non-Bayesian, non-hierarchical fitness plot, based on Table~S6 from Addinall et al. (2011) $(F=MDR\times MDP)$.
B) Non-Bayesian, hierarchical fitness plot, \hl{from fitting REM to data} in Table~S6 from Addinall et al. (2011) $(F=MDR\times MDP)$.
C) IHM fitness plot with $\emph{orf}\Delta$ posterior mean fitness $(F=MDR\times MDP)$.
D) JHM fitness plot with $\emph{orf}\Delta$ posterior mean fitnesses.
$\emph{orf}\Delta$ strains are classified as being a suppressor or enhancer based on analysis of growth parameter $r$.
Further explanation and notation for fitness plots are given in Figure~3 of the main article.
}
\end{figure}

\clearpage

\section{\label{app:interactions}Lists of top genetic interactions for IHM and JHM approaches}
\setcounter{figure}{0}
\setcounter{table}{0}
\begin{table}[h!]
\caption{Sample of IHM top genetic interactions\label{app:IHM_interactions}}
\centering
\resizebox{12cm}{!}{%
\npdecimalsign{.}
\nprounddigits{2}
\begin{tabular}{ c c n{2}{2} n{2}{2} c}
\hline
\emph{Type of}  & \emph{Gene} & \multicolumn{1}{c}{\emph{Probability of}} & \multicolumn{1}{c}{\emph{Strength of}}& \emph{Position in} \\ \emph{Interaction} & \emph{Name}  & \multicolumn{1}{c}{\emph{Interaction} ${\delta}_{l}$} & \multicolumn{1}{c}{\emph{Interaction} \emph{$e^{({\delta}_{l}{\gamma}_{l})}$}} & \emph{Addinall (2011)} \\ 
\hline
Suppressor & IPK1 & 1.00 & 2.87 & 10 \\ 
   & LST4 & 1.00 & 2.77 & 13 \\ 
   & RPN4 & 1.00 & 2.76 & 17 \\ 
   & MTC5 & 1.00 & 2.66 & 20 \\ 
   & GTR1 & 1.00 & 2.64 & 38 \\ 
   & NMD2 & 1.00 & 2.62 & 3 \\ 
   & SAN1 & 1.00 & 2.62 & 16 \\ 
   & UPF3 & 1.00 & 2.58 & 21 \\ 
   & RPL37A & 1.00 & 2.56 & 121 \\ 
   & NAM7 & 1.00 & 2.53 & 22 \\ 
   & RPP2B & 1.00 & 2.52 & 120 \\ 
   & YNL226W & 0.99 & 2.49 & 126 \\ 
   & YGL218W & 1.00 & 2.46 & 250 \\ 
   & MEH1 & 1.00 & 2.45 & 45 \\ 
   & ARO2 & 1.00 & 2.45 & 68 \\ 
   & EXO1 & 1.00 & 2.45 & 1 \\ 
   & BUD27 & 1.00 & 2.43 & 46 \\ 
   & RAD24 & 1.00 & 2.39 & 4 \\ 
   & RPL16B & 1.00 & 2.39 & 33 \\ 
   & RPL43A & 1.00 & 2.39 & 150 \\ 
\hline\noalign{\vskip 0.5mm} 
Enhancer  & MRC1 & 1.00 & 0.11 & 35 \\ 
   & YKU70 & 1.00 & 0.11 & 31 \\ 
   & STI1 & 1.00 & 0.11 & 42 \\ 
   & RIF1 & 1.00 & 0.13 & 36 \\ 
   & ELP3 & 1.00 & 0.16 & 82 \\ 
   & CLB5 & 1.00 & 0.17 & 58 \\ 
   & MRC1 & 1.00 & 0.17 & 63 \\ 
   & DPH2 & 1.00 & 0.18 & 24 \\ 
   & POL32 & 1.00 & 0.19 & 113 \\ 
   & MAK31 & 1.00 & 0.19 & 37 \\ 
   & SWM1 & 1.00 & 0.20 & 25 \\ 
   & LTE1 & 1.00 & 0.21 & 48 \\ 
   & MAK10 & 1.00 & 0.22 & 44 \\ 
   & ELP2 & 1.00 & 0.22 & 77 \\ 
   & PAT1 & 1.00 & 0.24 & 144 \\ 
   & DPH1 & 1.00 & 0.25 & 55 \\ 
   & SRB2 & 0.99 & 0.25 & 174 \\ 
   & THP2 & 1.00 & 0.26 & 67 \\ 
   & MFT1 & 1.00 & 0.26 & 52 \\ 
   & LSM6 & 0.97 & 0.26 & 389 \\ 
\hline
\multicolumn{5}{c}{A file containing the full list of genetic interactions is also provided in the on-line supporting materials.}
\end{tabular}
\npnoround
}
\end{table}

\begin{table}[h!]
\caption{Sample of JHM top genetic interactions\label{app:JHM_interactions}}
\centering
\resizebox{\columnwidth}{!}{%
\npdecimalsign{.}
\nprounddigits{2}
\begin{tabular}{ c c n{2}{2} n{2}{2} n{2}{2} n{2}{2} c}
\hline
\emph{Type of}  & \emph{Gene} & \multicolumn{1}{c}{\emph{Probability of}} & \multicolumn{1}{c}{\emph{Strength of}} & \multicolumn{1}{c}{\emph{Strength of}} & \emph{Strength of} & \multicolumn{1}{c}{\emph{Position in}} \\ 
 \emph{Interaction} & \emph{Name} & \multicolumn{1}{c}{\emph{Interaction}} & \multicolumn{1}{c}{\emph{Interaction}} & \multicolumn{1}{c}{\emph{Interaction}} & \emph{Interaction} & \emph{Addinall (2011)} \\ 
 & &\multicolumn{1}{c}{${\delta}_l$} & \multicolumn{1}{c}{ \emph{$e^{({\delta}_{l}{\gamma}_l)}$}} & \multicolumn{1}{c}{\emph{$e^{({\delta}_{l}{\omega}_{l})}$}} & \emph{$MDR \times MDP$} & \\
\hline
Suppressor   & CSE2 & 1.00 & 490.51 & 0.48 & 11.71 & 838 \\ 
in K   & SGF29 & 1.00 & 273.69 & 0.68 & 14.16 & 580 \\ 
   & GSH1 & 1.00 & 78.79 & 0.92 & 17.89 & 281 \\ 
   & YMD8 & 1.00 & 59.31 & 0.65 & 7.05 & 2022 \\ 
   & YGL024W & 1.00 & 28.13 & 1.18 & 13.33 & 151 \\ 
   & RPS9B & 1.00 & 24.67 & 1.12 & 10.24 & 801 \\ 
   & GRR1 & 1.00 & 22.51 & 0.67 & 5.99 & 1992 \\ 
\hline\noalign{\vskip 0.5mm} 
 
Suppressor   & BTS1 & 1.00 & 19.27 & 2.29 & 19.65 & 201 \\ 
in r   & IPK1 & 1.00 & 5.56 & 2.26 & 44.81 & 10 \\ 
   & NMD2 & 1.00 & 2.96 & 2.19 & 48.51 & 3 \\ 
   & SAN1 & 1.00 & 2.37 & 2.17 & 48.70 & 16 \\ 
   & LST4 & 1.00 & 5.79 & 2.14 & 44.14 & 13 \\ 
   & RPN4 & 1.00 & 8.00 & 2.12 & 40.46 & 17 \\ 
   & UPF3 & 1.00 & 3.16 & 2.07 & 45.25 & 21 \\ 

\hline\noalign{\vskip 0.5mm} 
Suppressor in   & SAN1 & 1.00 & 2.37 & 2.17 & 48.70 & 16 \\ 
 $MDR\times MDP$  & NMD2 & 1.00 & 2.96 & 2.19 & 48.51 & 3 \\ 
   & UPF3 & 1.00 & 3.16 & 2.07 & 45.25 & 21 \\ 
   & EXO1 & 1.00 & 2.89 & 2.06 & 45.04 & 1 \\ 
   & IPK1 & 1.00 & 5.56 & 2.26 & 44.81 & 10 \\ 
   & LST4 & 1.00 & 5.79 & 2.14 & 44.14 & 13 \\ 
   & NAM7 & 1.00 & 3.02 & 2.04 & 43.00 & 22 \\
\hline\noalign{\vskip 0.5mm} 
Enhancer   & YKU70 & 1.00 & 0.01 & 1.09 & -23.44 & 31 \\ 
 in K  & STI1 & 1.00 & 0.01 & 1.20 & -21.60 & 42 \\ 
   & RIF1 & 1.00 & 0.01 & 0.63 & -26.17 & 36 \\ 
   & MRC1 & 1.00 & 0.01 & 0.83 & -23.15 & 35 \\ 
   & MAK31 & 1.00 & 0.02 & 1.18 & -18.19 & 37 \\ 
   & CLB5 & 1.00 & 0.02 & 0.87 & -19.54 & 58 \\ 
   & MRC1 & 1.00 & 0.02 & 0.81 & -20.40 & 63 \\ 
\hline\noalign{\vskip 0.5mm} 
Enhancer   & PAT1 & 1.00 & 1.71 & 0.28 & -18.30 & 144 \\ 
 in r  & PUF4 & 1.00 & 2.00 & 0.31 & -21.61 & 34 \\ 
   & YKU80 & 1.00 & 2.15 & 0.33 & -21.68 & 32 \\ 
   & RTT103 & 1.00 & 2.54 & 0.34 & -17.87 & 153 \\ 
   & LSM1 & 0.99 & 2.13 & 0.34 & -16.20 & 101 \\ 
   & GIM3 & 0.99 & 0.93 & 0.35 & -19.70 & 132 \\ 
   & INP52 & 0.96 & 0.86 & 0.36 & -14.50 & 345 \\ 

\hline\noalign{\vskip 0.5mm} 

Enhancer in   & RIF1 & 1.00 & 0.01 & 0.63 & -26.17 & 36 \\ 
  $MDR\times MDP$  & LTE1 & 1.00 & 0.06 & 0.40 & -23.96 & 48 \\ 
   & YKU70 & 1.00 & 0.01 & 1.09 & -23.44 & 31 \\ 
   & MRC1 & 1.00 & 0.01 & 0.83 & -23.15 & 35 \\ 
   & DPH2 & 1.00 & 0.04 & 0.56 & -23.11 & 24 \\ 
   & EST1 & 1.00 & 0.12 & 0.46 & -22.20 & 5 \\ 
   & MAK10 & 1.00 & 0.04 & 0.59 & -21.92 & 44 \\ 

\hline
\multicolumn{7}{c}{A file containing the full list of genetic interactions is also provided in the on-line supporting materials.}
\end{tabular}
\npnoround
}
\end{table}
 
\clearpage

\section{\label{app:alternative_fit}Alternative fitness plots for the JHM}
\setcounter{figure}{0}
\setcounter{table}{0}
\begin{figure}[h!]
  \centering
\includegraphics[width=14cm]{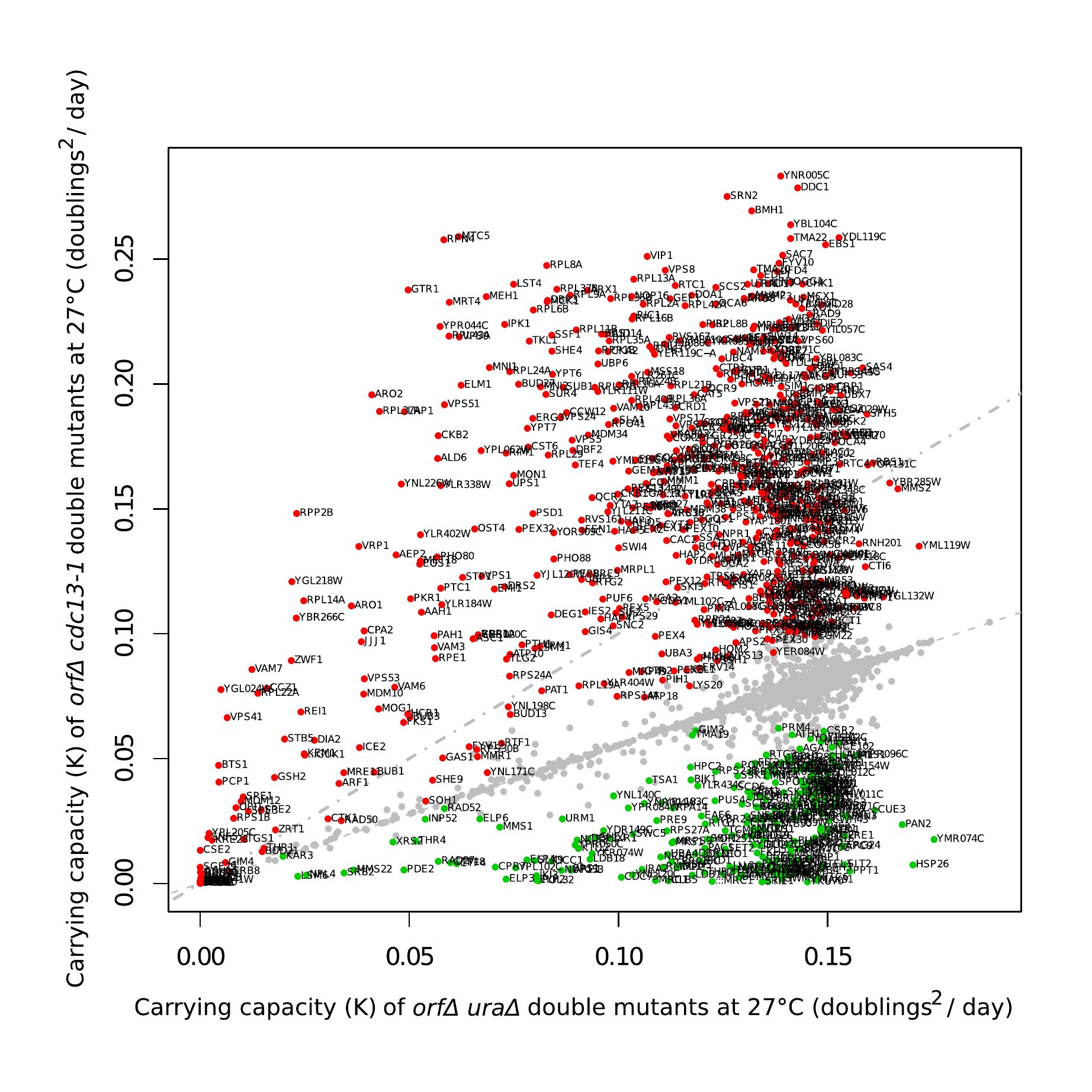}
\caption{Joint hierarchical model (JHM) carrying capacity fitness plot with $\emph{orf}\Delta$ posterior mean fitnesses.
$\emph{orf}\Delta$ strains are classified as being a suppressor or enhancer based on carrying capacity parameter $K$.
Further explanation and notation for fitness plots are given in Figure~3 of the main article.\label{fig:JHM_K}
}
\end{figure}

\begin{figure}[h!]
  \centering
\includegraphics[width=14cm]{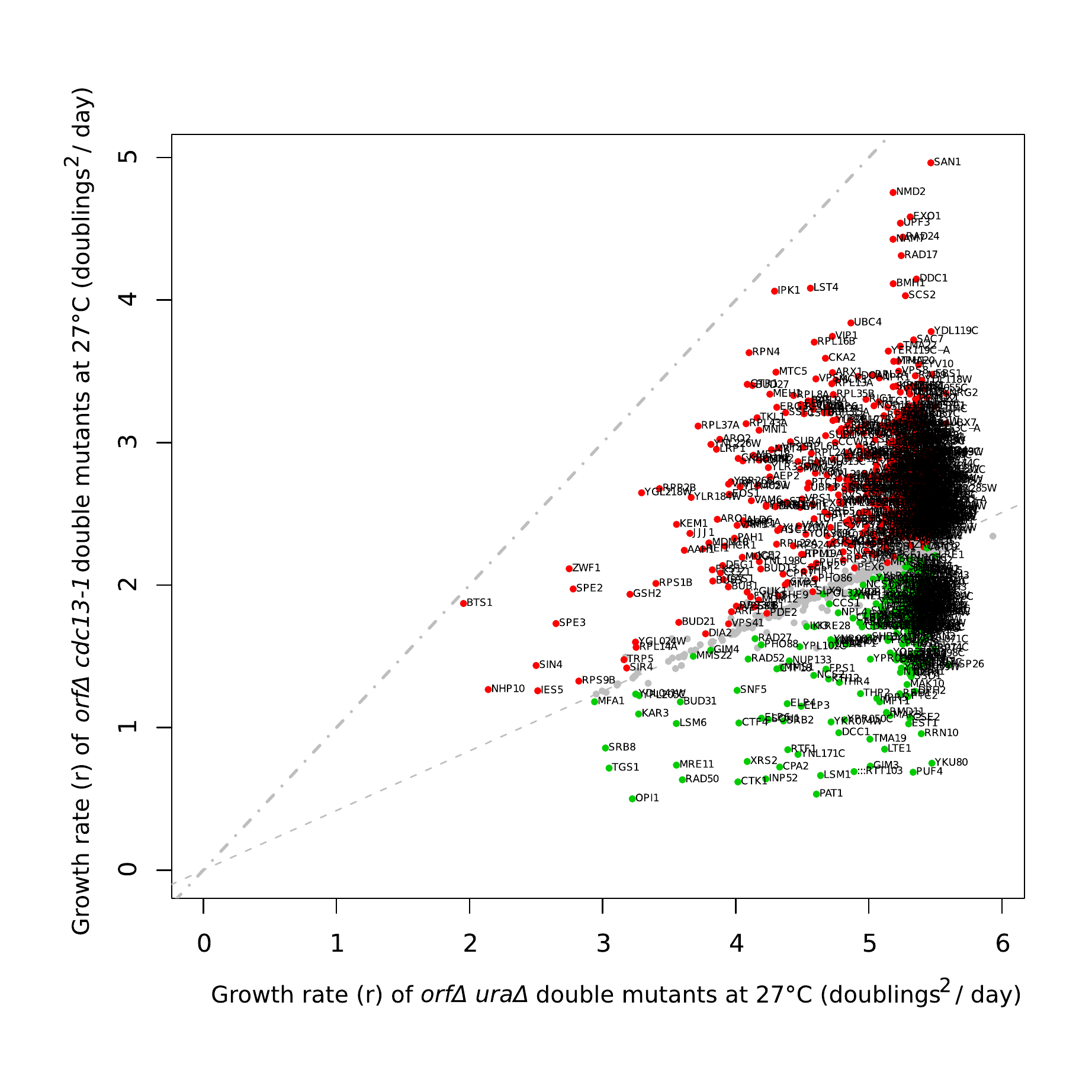}
\caption{Joint hierarchical model (JHM) growth rate fitness plot with $\emph{orf}\Delta$ posterior mean fitnesses.
$\emph{orf}\Delta$ strains are classified as being a suppressor or enhancer based on growth parameter $r$.
Further explanation and notation for fitness plots are given in Figure~3 of the main article.\label{fig:JHM_r}
}
\end{figure}

\clearpage

\section{The effect of parameter $p$ on specificity}

\setcounter{figure}{0}
\setcounter{table}{0}
\begin{figure}[h!]
  \centering
\includegraphics[width=12cm]{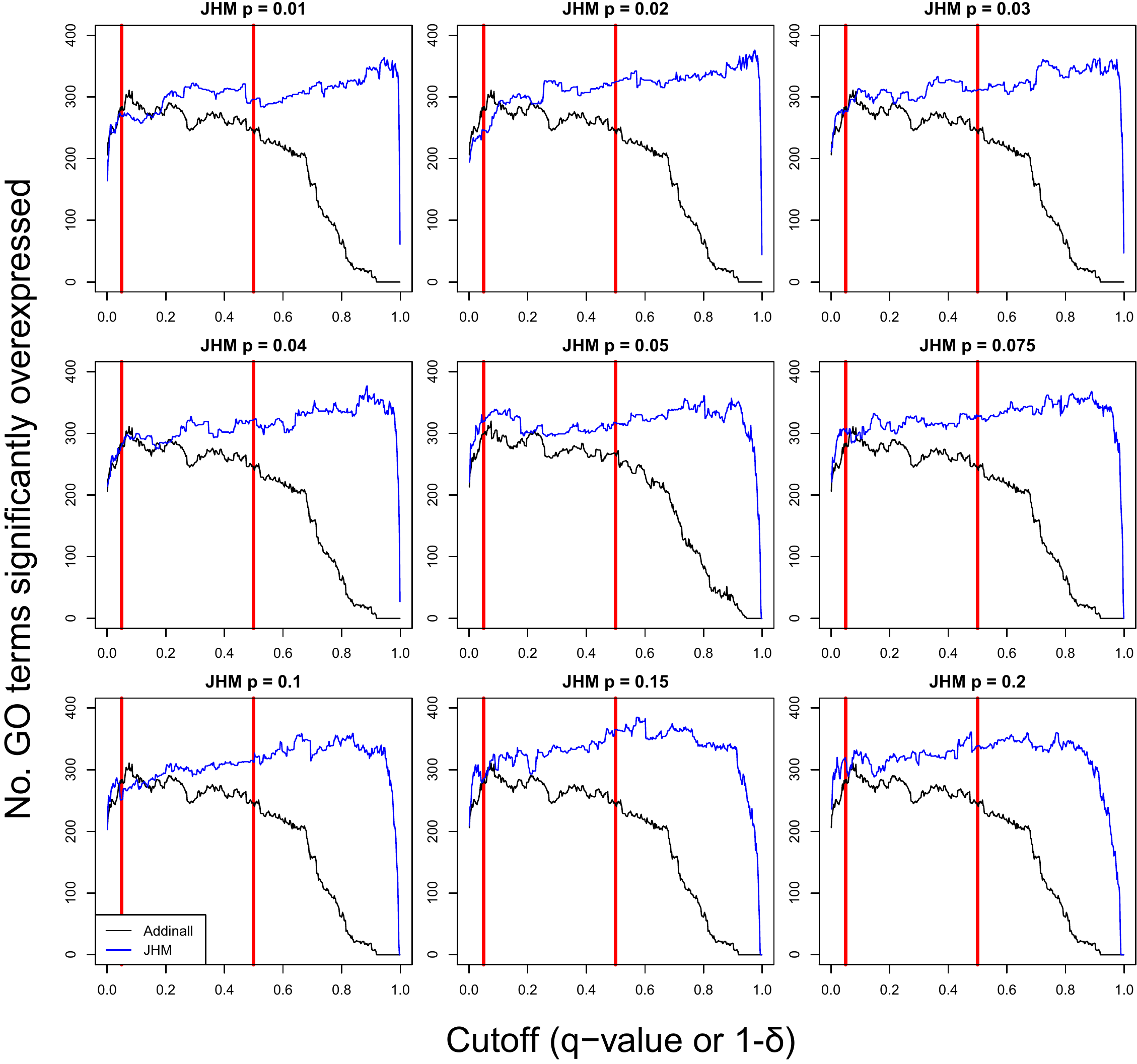}
\caption{ 
Comparison of the number of significantly over-expressed GO terms identified in lists of significant interactors found using the Addinall (2011) method and using the JHM.  Significantly over-expressed GO terms were identified using the hyperGTest function in the GOstats R package.  Note that the values used to classify whether a gene interacts with \emph{cdc13-1} at 27C (q-value and $\delta$ respectively, red vertical lines as presented in Section~4.4) are not directly comparable.
However, the full range of possible cutoffs for both values are plotted.  Each panel shows the change in over-expressed GO terms with cutoff for a different value of the $p$ parameter (prior estimate of expected proportion of interactors) used in the JHM analysis.
}
\label{app:sens_spec_supp}
\end{figure}

\clearpage

\section{Correlation between methods}
The Addinall et al. (2011) approach has its highest correlation with the IHM, followed by the JHM and then the REM. 
The REM correlates least well with the JHM while showing the same correlation with both the Addinall et al. (2011) approach and the IHM.
The correlation between the IHM and the JHM is the largest observed between any of the methods, demonstrating the similarity of our Bayesian hierarchical methods. 
\begin{table}[h!]
\caption{Spearman's rank correlation coefficients for magnitudes from genetic independence, between Addinall et al. (2011), REM, IHM and JHM QFA methods \label{tab:spearman}}
\centering 
\resizebox{\columnwidth}{!}{%
  \begin{tabular}{*{5}{c}}
    \\
  	   \hline
  \\
\emph{Method} &\multicolumn{4}{c}{\emph{Method}}\\ 
&&& \\ \cline{2-5}
&&&\\
   		  & \emph{Addinall et al. (2011)} & \emph{REM} & \emph{IHM} & \emph{JHM QFA} \\
			   		  & \emph{QFA} & \emph{QFA} & \emph{QFA} & \emph{($MDR\times MDP$)} \\	
   		  \\
   		   \hline
\\   		  
Addinall et al. (2011) QFA,   	&1	& 0.77 & 0.89 & 0.88 \\ 	

REM QFA,    					& 	&1	 & 0.77	& 0.75 \\

IHM QFA,   						&	  &    &1	& 0.95 \\

JHM QFA ($MDR\times MDP$),    			& 	&	   &  & 1 \\

\\		
\hline
  		 \end{tabular}
			}
\end{table}

The $MDR\times MDP$ correlation plot of the JHM versus the Addinall et al. (2011) approach demonstrates the similarity (Pearson correlation=0.90) and differences between the two approaches in terms of $MDR\times MDP$.
\rev{We can see how the results differ between the JHM and Addinall et al. (2011), with a kink at the origin due to the JHM allowing shrinkage of non-interacting genes towards the fitted line.}

\setcounter{figure}{0}
\setcounter{table}{0}
\clearpage
\begin{figure}[h!]
  \centering
\includegraphics[width=14cm]{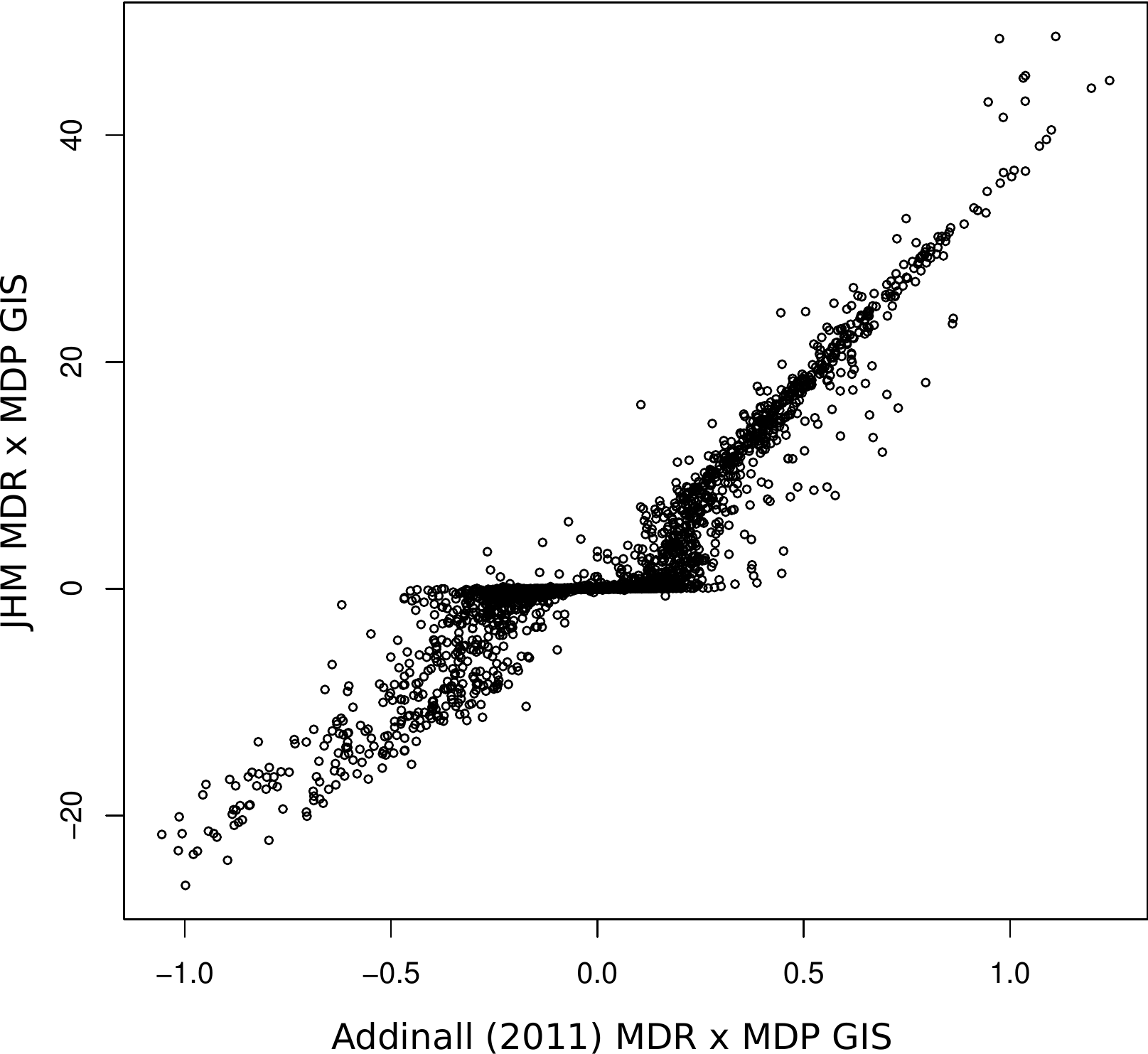}
\caption{ $MDR\times MDP$ genetic interaction correlation plot of JHM versus Addinall et al. (2011) (Pearson correlation=0.90).
}
\label{app:correlation_JHM_ADD}
\end{figure}

\clearpage

\section{\label{app:simstudy}A simulation study comparing specificity and sensitivity of the Addinall et al. (2011) approach, the SHM and the JHM}

Since our understanding of biological processes is currently incomplete it is difficult to assess what proportion of genetic interactions identified by fitting any model to real biological data are real.  If we don't know which interactions are true, we cannot know which of those interactions identified by any inference scheme are false positives.  In order to compare the ability of each of our models to identify subtle, true interactions (sensitivity) while avoiding false positives (specificity), a separate simulation study was carried out (Section 4.3.6 http://arxiv.org/abs/1405.7091).  Synthetic control and query datasets of similar size, quality and resolution to real QFA datasets, with known suppressors and enhancers of a simulated query mutation were constructed using a hierarchical simulation model consistent with the JHM.  We used the JHM to simulate the synthetic dataset since it is the most detailed model we have available and the one which most closely matches the structure of QFA experiments.  The Addinall et al. (2011) approach, the REM, the SHM and the JHM were each fit to the synthetic dataset and the lists of suppressors and enhancers as well as the list of all interactors generated by each method were compared with the list of known true interactors.

Sensitivity and specificity achieved with each of the models were presented in table 4.4 http://arxiv.org/abs/1405.7091.  In summary, the simulation study showed that the JHM correctly identified a higher proportion of true interactions (314/430) than the Addinall et al. (2011) approach (220/430), while also identifying fewer false positives (JHM: 8, Addinall et al. (2011): 303).